\documentclass[iop,revtex4,onecolumn]{emulateapj}

\slugcomment{Accepted by ApJ}

\shorttitle{Radio counts of star-forming galaxies}
\shortauthors{C. Mancuso et al.}

\usepackage{relsize}
\usepackage{graphicx}
\usepackage{epstopdf}
\usepackage{rotating}
\usepackage{lscape}
\usepackage{amsmath,amssymb}
\usepackage{natbib}
\everymath{\displaystyle}
\DeclareMathAlphabet{\mathpzc}{OT1}{pzc}{m}{it}

\newcommand{\angstrom}{{\rm \mathring A}}

\def\lsim{\,\lower2truept\hbox{${<\atop\hbox{\raise2truept\hbox{$\sim$}}}$}\,}
\def\gsim{\,\lower2truept\hbox{${>\atop\hbox{\raise2truept\hbox{$\sim$}}}$}\,}

\begin{document}
\title{Predictions for ultra-deep radio counts of star-forming galaxies}
\author{Claudia Mancuso\altaffilmark{1}, Andrea Lapi\altaffilmark{2,1}, Zhen-Yi Cai\altaffilmark{3}, Mattia Negrello\altaffilmark{4}, Gianfranco De Zotti\altaffilmark{4,1}, Alessandro Bressan\altaffilmark{1}, Matteo Bonato\altaffilmark{4}, Francesca Perrotta\altaffilmark{1}, Luigi Danese\altaffilmark{1}}
\altaffiltext{1}{Astrophysics Sector, SISSA, Via Bonomea 265, I-34136
Trieste, Italy; cmancuso@sissa.it} \altaffiltext{2}{Dipartimento di Fisica,
Universit\`a `Tor Vergata', Via della Ricerca Scientifica 1, I-00133 Roma,
Italy} \altaffiltext{3}{CAS Key Laboratory for Research in Galaxies and Cosmology, Department of Astronomy, University of Science and Technology of China, Hefei, Anhui 230026, China}
\altaffiltext{4}{INAF - Osservatorio Astronomico di Padova, Vicolo
dell'Osservatorio 5, I-35122 Padova, Italy}

\def\LaTeX{L\kern-.36em\raise.3ex\hbox{a}\kern-.15em
    T\kern-.1667em\lower.7ex\hbox{E}\kern-.125emX}
\def\simlt{\mathrel{\rlap{\lower 3pt\hbox{$\sim$}}\raise 2.0pt\hbox{$<$}}}
\def\simgt{\mathrel{\rlap{\lower 3pt\hbox{$\sim$}}\raise 2.0pt\hbox{$>$}}}

\begin{abstract}
We have worked out predictions for the radio counts of star-forming
galaxies down to nJy levels, along with redshift distributions down to the
detection limits of the phase 1 Square Kilometer Array MID telescope
(SKA1-MID) and of its precursors. Such predictions were obtained by
coupling epoch dependent star formation rate (SFR) functions with relations
between SFR and radio (synchrotron and free-free) emission. The SFR
functions were derived taking into account both the dust obscured and the
unobscured star-formation, by combining far-infrared (FIR), ultra-violet
(UV) and H$\alpha$ luminosity functions up to high redshifts. We have also
revisited the South Pole Telescope (SPT) counts of dusty galaxies at
95\,GHz performing a detailed analysis of the Spectral Energy Distributions
(SEDs). Our results show that the deepest SKA1-MID surveys will detect
high-$z$ galaxies with SFRs two orders of magnitude lower compared to
\textit{Herschel} surveys. The highest redshift tails of the distributions
at the detection limits of planned SKA1-MID surveys comprise a substantial
fraction of strongly lensed galaxies. We predict that a survey down to
$0.25\,\mu$Jy at 1.4 GHz will detect about 1200 strongly lensed galaxies
per square degree, at redshifts of up to 10. For about  30\% of them the
SKA1-MID will detect at least 2 images. The SKA1-MID will thus provide a
comprehensive view of the star formation history throughout the
re-ionization epoch, unaffected by dust extinction. We have also provided
specific predictions for the EMU/ASKAP and MIGHTEE/MeerKAT surveys.
\end{abstract}


\section{Introduction}\label{sect:intro}

There are several well established diagnostic tools for measuring the
star formation rates (SFRs) in galaxies, each with its own strengths and
weaknesses \citep[see][for a comprehensive review]{KennicuttEvans2012}. The
UV and the H$\alpha$ luminosities are among the best direct tracers of the
emission from young stars, so that the SFR is simply proportional to the
luminosity. These diagnostics, however, are highly sensitive to dust
extinction and may miss obscured star formation. The latter is measured by
the far-infrared/sub-mm luminosity due to dust reprocessing of the emission
from newly formed stars. But the dust heating is not always dominated by
young stars: older stellar populations can contribute as well. Both the
UV/H$\alpha$ and the far-infrared (FIR) diagnostics may also be contaminated
by emissions from Active Galactic Nuclei (AGNs).

Ever since radio surveys have reached sub-mJy flux density levels
\citep{Windhorst1985}, they proved to be a primary means of identifying
star-forming galaxies at high redshift \citep[e.g.,][]{Bonzini2013}.
Observations have solidly demonstrated a tight correlation between the
low-frequency radio continuum, due to synchrotron emission of relativistic
electrons mostly  produced by supernovae, and far-IR emission of galaxies,
legitimating its application as a SFR tracer
\citep[see][]{Helou1985,Condon1992,Yun2001,Ivison2010,Jarvis2010,Bourne2011,Mao2011}.
A word of caution is in order, however, because the physical basis of this
relation is not fully understood.  Actually there are many physical
processes, such as propagation of relativistic electrons, strength and
structure of the magnetic field, size and composition of dust grains, that
must cooperate to produce this relation \citep{Bell2003,Lacki2010}.
As a consequence it is not granted that this relation applies also to
redshift/luminosity ranges where the available data are insufficient to test
it precisely. In addition the synchrotron emission could be contaminated,
even by a large factor, by faint nuclear radio activity, and this would bias
the estimation of the SFR.

A more direct radio SFR tracer is the free-free emission from the gas ionized
by massive young stars \citep{Murphy2009,Murphy2015}. Being directly
proportional to the production rate of ionising photons it provides a measure
of the SFR without the complication of dust attenuation encountered in the
optical/UV. It has a flat spectrum and is expected to show up at frequencies
of tens of GHz. The interpretation of data at these frequencies may be
complicated by the presence of ``anomalous'' dust emission \citep[][and
references therein]{PlanckCollaborationXX2011} attributed to spinning dust
grains \citep[e.g.,][]{DraineLazarian1998}. However a significant
contribution of this component to the global emission of galaxies has not
been proved yet \citep{Murphy2012,PlanckCollaborationXXV2014}.

The current deepest radio surveys \citep[see][for a review]{DeZotti2010} have
not been carried out at frequencies high enough or are not deep enough to see
the transition from the synchrotron to the free-free dominance.  Only with
the advent of the Square Kilometer Array (SKA), and partially with its
precursors, we expect to efficiently select high-$z$ star-forming galaxies
via their free-free emission.

In this paper we carry out a thorough investigation of the radio counts and
redshift distributions of star forming galaxies, down to the flux density
levels that can be reached by the SKA. Our predictions rely on a
state-of-the-art model \citep{Cai2014} that accounts for the evolution of the
cosmic SFR function up to $z\simeq 10$ as measured by a combination of
far-infrared (FIR), UV and H$\alpha$ data. An outline of the model is
presented in Sect.~\ref{sect:model}. The redshift dependent SFR function is
translated into the evolving radio luminosity function, taking into account
both synchrotron and free-free emissions, using the calibrations by
\citet{Murphy2011}. In Sect.~\ref{sect:calibration} we tune the correlation
between the SFR and the synchrotron luminosity in order to match the
observational determination of the local luminosity function of star-forming
galaxies \citep{MauchSadler2007}. This allows us to reproduce the available
multi-frequency source counts. Those of star-forming galaxies at 95 GHz
provided by the South Pole Telescope (SPT) survey \citep{Mocanu2013}, that
are the most sensitive to the free-free emission, have been reassessed
performing a more accurate discrimination between star-forming galaxies and
radio-loud active galactic nuclei.  Having the radio luminosity function at
each redshift it is straightforward to work out predictions for counts and
redshift distributions of star-forming galaxies. In
Sect.~\ref{sect:SKAcounts} we show a selection of the results for the
frequency range 1.4 -- 30 GHz and compare the coverage of the SFR--$z$ plane
by \textit{Herschel}, optical/UV surveys and the phase 1 SKA-MID.
Section~\ref{sect:synergies} explores possible synergies with deep
optical/near-infrared (near-IR) surveys.  Finally,
Sect.~\ref{sect:conclusions} summarizes our main conclusions.

Throughout this paper we adopt a flat $\Lambda \rm CDM$ cosmology with matter
density $\Omega_{\rm m} = 0.32$, $\Omega_{\rm b} = 0.049$, $\Omega_{\Lambda}
= 0.68$, Hubble constant $h=H_0/100\, \rm km\,s^{-1}\,Mpc^{-1} = 0.67$,
spectrum of primordial perturbations with index $n = 0.96$ and normalization
$\sigma_8 = 0.83$ \citep{PlanckCollaborationXVI2013}.

\section{Outline of the model}\label{sect:model}

As illustrated by the right-hand panel of Fig.~8 of
\citet{MadauDickinson2014} and by Fig.~10 of \citet{Cai2014}, the
\textit{mean} dust attenuation is low at $z\simgt 6$, when the metallicity of
the intergalactic medium (IGM) is lower than the solar value, $Z_\odot$, by a
factor $\simlt 10^{-3}$ \citep[Fig.~14 of][]{MadauDickinson2014}.
Thus, at these redshifts, measurements at UV wavelengths are likely
to be a reliable SFR diagnostic for most galaxies, although a few galaxies with
high SFRs and substantial dust extinction have been found
\citep{Riechers2013,Watson2015}. Great progress has been made in recent
years in the determination of UV luminosity functions at high $z$, up to
$z\sim 10$
\citep{Bouwens2008,Bouwens2011a,Bouwens2011b,Bouwens2015,Smit2012,Oesch2012,Oesch2013a,Oesch2013b,Schenker2013,McLure2013},
struggling to reconstruct the history of the cosmic re-ionization.

The mean dust attenuation increases rapidly with decreasing $z$ down to
$z\sim 1$--1.5, where it peaks, to decline at still lower $z$. In the range
$1\simlt z \simlt 3$, including the peak in the cosmic SFR density, the star
formation is heavily dust enshrouded, so that the IR (8--$1000\,\mu$m)
luminosity is an accurate SFR measure. At the other redshifts the safest
approach combines UV with IR measurements. In dealing with the latter, we
need to be aware that, especially at low redshifts, a significant
contribution to dust heating may be due to older stellar populations.

A detailed study of the evolution of the SFR function across the cosmic
history was carried out by \citet{Cai2013} focussing on IR data, and by
\citet{Cai2014} at $z\simgt 2$ focussing on UV and Ly$\alpha$ data but taking
into account also dust attenuation and re-emission. These studies build on
the work by \citet{Granato2004}, further developed by
\citet{Lapi2006,Lapi2011,Lapi2014}, \citet{Mao2007}. The underlying scenario
relies on the results of  intensive $N$-body simulations and semi-analytic
studies \citep[e.g.,][]{Zhao2003,Wang2011,LapiCavaliere2011} showing that
pre-galactic halos undergo an early phase of fast collapse, including a few
major merger events, during which the central regions reach rapidly a
dynamical quasi-equilibrium. A slow growth of the halo outskirts in the form
of many minor mergers and diffuse accretion follows. This second stage has
little effect on the inner potential well where the visible galaxy resides.

The baryons falling into the potential wells created during the fast collapse
phase are shock-heated to the virial temperature. The gas cools down giving
rise to star formation. At the same time it loses angular momentum, e.g. by
effect of the radiation drag, and flows towards the central regions,
accreting onto the central super-massive black hole. These processes are
governed by a set of equations, reported in the Appendix of \citet{Cai2013},
that take into account the energy feedback from supernovae and from the
active nucleus. The model includes a self consistent treatment of the
chemical evolution of the ISM, calculated using the standard equations and
stellar nucleosynthesis prescriptions, as written down, for instance, in
\citet{Romano2002}. The chemical evolution controls the evolution of the dust
abundance, hence the dust absorption and re-emission.

In this framework \citet{Cai2014} worked out a model for the evolution of the
ultraviolet (UV) and Ly$\alpha$ luminosity functions of high-redshift
($z\simgt 2$) galaxies, taking into account their chemical evolution and the
associated evolution of dust extinction. The model yields good fits of the
observed luminosity functions at all redshifts at which they have been
measured and provides a physical explanation for the weak observed evolution
of both luminosity functions between $z =2$ and $z = 6$.

While at high $z$ the high star formation efficiency implied by the scenario
briefly sketched above applies to most galaxies, at lower $z$ a dichotomy
arises \citep[see, e.g.,][]{Bernardi2010}. Early-type galaxies and massive
bulges of Sa galaxies are composed of relatively old stellar populations with
mass-weighted ages $\simgt 8$--9\,Gyr. They must therefore have been forming
most of their stars over short timescales at $z\simgt 1$--1.5 and must have
dominated the cosmic SFR at high $z$. Instead, the disc components of spirals
and the irregular galaxies are characterized by significantly younger stellar
populations  and star formation activity continuing up to the present time.
They are therefore the dominant star formers at $z\simlt 1.5$.

\citet{Cai2013} modeled the cosmological evolution of the SFR function of
proto-spheroidal galaxies (the progenitors of present day early-type galaxies
and of massive bulges of disc galaxies) during their dust-enshrouded
star-formation phase, and the concomitant evolution of their active nuclei.
This model provides a physical explanation for the observed positive
evolution of both galaxies and AGNs up to $z\approx2.5$ and for the negative
evolution at higher redshifts, for the sharp transition from Euclidean to
extremely steep counts at (sub-)mm wavelengths, as well as for the (sub-)mm
counts of strongly lensed galaxies, that are hard to account for by
alternative, physical or phenomenological, approaches. Furthermore, as shown
by \citet{Xia2012} and \citet{Cai2013}, the halo masses inferred both from
the angular correlation function of detected sub-mm galaxies
\citep{Cooray2010,Maddox2010} and from the power spectrum of fluctuations of
the cosmic infrared background
\citep{Amblard2011,PlanckCollaborationXVIII2011,PlanckCollaborationXXX2013,Viero2013}
are fully consistent with those implied by this scenario while they are
larger than those implied by the major mergers plus top-heavy initial stellar
mass function \citep{Baugh2005,Lacey2010} and smaller than those implied by
cold flow models \citep{Dave2010}.

Late-type galaxies follow a different evolutionary track.  The evolution of
the IR luminosity function of these objects was described by \citet{Cai2013}
using a phenomenological approach, considering two populations with different
spectral energy distributions (SEDs) and different evolutionary properties:
``normal'' late-type galaxies, with low evolution and low dust temperatures
and rapidly evolving starburst galaxies, with warmer dust temperatures.

The evolution of late-type ``normal'' and ``starburst'' galaxies is described
using a phenomenological parametric approach. For the starburst population,
power-law density and luminosity evolution
[$\Phi^{\star}(z)=\Phi^{\star}(z=0)\times(1+z)^{\alpha_{\Phi}}$;
$L^{\star}(z)=L^{\star}(z=0)\times(1+z)^{\alpha_{L}}$] up to $z_{\rm
break}=1$ was assumed. The normal galaxy population has only a mild
luminosity evolution up to the same value of $z_{\rm break}$, as indicated by
chemo/spectrophotometric evolution models.  At $z>z_{\rm break}$ both
$\Phi^{\star}(z)$ and $L^{\star}(z)$ are kept to the values at $z_{\rm
break}$ multiplied by a smooth cut-off function. For further details and the
values of the parameters, see \citet{Cai2013}.

The model fits a broad variety of data\footnote{See
http://people.sissa.it/$\sim$zcai/galaxy\_agn/ or
http://staff.ustc.edu.cn/$\sim$zcai/galaxy\_agn/index.html.}: multi-frequency
and multi-epoch luminosity functions of galaxies and AGNs, redshift
distributions, number counts (total and per redshift bins). It also
accurately accounts for the recent  counts and redshift distribution of
strongly lensed galaxies detected by the South Pole Telescope
\citep[SPT;][]{Mocanu2013,Weiss2013} as shown by \citet{Bonato2014}.

When the dust heating is dominated by young stars and the effective
optical/UV optical depth is high, the total (8--$1000\,\mu$m) IR luminosity,
$L_{\rm IR}$, is a good measure of the SFR. More generally, the IR luminosity
is the sum of a ``warm'' dust component heated by young stars and of a
``cold'' dust (or ``cirrus'') component, heated by the general radiation
field probably dominated by older stars. Obviously the only tracer of the SFR
is the first component. This issue was investigated by \citet{Clemens2013}
using a complete sample of local star-forming galaxies detected by the
\textit{Planck} satellite. We adopt the relation between SFR and $L_{\rm IR}$
derived by these authors:
\begin{equation}\label{eq:LIR}
\log({\rm SFR}/\rm M_{\odot}\,yr^{-1}) = -9.6 + \log(L_{\rm IR}/L_\odot) -
\bigg(\frac{2.0}{\log(L_{\rm IR}/L_\odot) -7.0}\bigg),
\end{equation}
As expected, high IR luminosities generally correspond to high SFRs, i.e. to
a dominant ``warm'' dust component. Note that calibrations of SFR diagnostics
generally depend on the adopted initial mass function (IMF). In the above
equation we have adopted the one by \citet{Chabrier2003}. According to Fig.~1
of \citet{ChomiukPovich2011} the differences with the calibrations using the
\citet{Kroupa2001} IMF \citep{Murphy2011,Murphy2012,KennicuttEvans2012} are
$\lesssim$ 10$\%$ and will be neglected. The \citet{Cai2013} model deals only
with dust-obscured star formation. To get a comprehensive view of the
evolution of the SFR function of late-type galaxies, including the unobscured
star formation, we have complemented it taking into account optical/UV SFR
indicators, adopting again a parametric approach for
evolution\footnote{Early-type galaxies are in passive evolution at low $z$
and their early unobscured star-formation phases are dealt with by
\citet{Cai2014}}. Details are given in the Appendix~\ref{appA}.

The SFR functions at several redshifts, obtained combining IR, UV and
H$\alpha$ survey data, are shown in Fig.~\ref{fig:SFRF}, where also the SFR
functions derived from the UV luminosity functions using the
\citet{KennicuttEvans2012} calibration but without the extinction correction
are plotted (thin solid lines). Below a $\hbox{SFR}\simeq
0.1$--$1\,M_\odot\,\hbox{yr}^{-1}$, depending on $z$, the SFR functions
derived from the IR luminosity functions fall below those derived from the UV
and H$\alpha$ luminosity functions without extinction corrections, implying
that extinction becomes negligible for low SFRs. The opposite happens for
higher SFRs. The model tends to underpredict the brightest points of
the luminosity functions. We note, however, that the discrepancy is of the
order of differences among estimates from different surveys. This suggests
that the data points are affected by cosmic variance due to the relatively
small area studied; the rare objects with the highest luminosities are
preferentially detected in areas where they are overdense.

While at $z< 1.5$ the observationally determined SFR functions are fully
accounted for by UV measurements with the appropriate extinction corrections,
at higher redshifts UV surveys miss most of the galaxies with the highest
SFRs, that are heavily dust obscured and therefore show up at far-IR/sub-mm
wavelengths. According to our model, the dominant contribution at the highest
SFRs at $z\simgt 1.5$ comes from proto-spheroidal galaxies.

Figure~\ref{fig:SFRevol} presents a synoptic view of the cosmological
evolution of the cosmic SFR function up to $z=8$ as given by the model
described above, including the contributions of normal late-type, starburst
and proto-spheroidal galaxies. This function has a complex evolutionary
behaviour. Its knee value increases with $z$ up to $z\simeq 2$--3 and
declines at higher redshifts; its slopes above and below the knee also change
with redshift. \citet{Aversa2015} have worked out approximate analytic
expressions providing a sensible rendition of the observational
determinations of the redshift-dependent SFR function. To this end they used
a modified Schechter function with evolving characteristic luminosity and
slopes. The luminosity function [$\hbox{Mpc}^{-3}\,\log({\rm SFR})^{-1}$]
writes
\begin{equation}\label{eq:SFRfun}
\phi_{\rm SFR}(z) = \Phi(z)\, \left[{\rm SFR}\over {\rm
SFR}_\star(z)\right]^{1-\alpha(z)}\,\exp{\left\{-\left[{\rm SFR}\over {\rm
SFR}_\star(z)\right]^{\omega(z)}\right\}},
\end{equation}
where
\begin{eqnarray}\label{eq:SFRfun_par}
\log \Phi(z)\, [\mathrm{Mpc}^{-3}\, \log({\rm SFR})^{-1}] &=&  -2.4 -2.3\, \chi +6.2\, \chi^2 -4.9\, \chi^3,\nonumber \\
\log \hbox{SFR}_\star(z)\, [M_\odot~\mathrm{yr}^{-1}] &=&  1.1 +3.2\, \chi -1.4\, \chi^2 - 2.1\, \chi^3,\nonumber \\
\alpha(z) &=&  1.2 + 0.5\, \chi -0.5\, \chi^2 +0.2\, \chi^3,\nonumber \\
\omega(z) &=&  0.7 -0.15\, \chi +0.16\, \chi^2 +0.01\, \chi^3,\nonumber
\end{eqnarray}
with $\chi= \log (1+z)$. Note that the above equation is simply meant
to provide an analytic rendition of the complex evolution of the SFR
function, for use in practical applications. As pointed out by
\citet{Aversa2015} the non-homogeneous nature of the data sets used to derive
the SFR functions and the diverse systematics affecting them do not warrant
the use of formal fitting procedures.

\section{Relations between radio emission and SFR}\label{sect:calibration}

\subsection{Calibration of the relation between SFR and synchrotron emission}\label{sect:sync}

A calibration of the SFR-synchrotron luminosity relation was obtained by
\citet{Murphy2011} using Starburst99 \citep{Leitherer1999} with a
\citet{Kroupa2001} IMF and specific choices for the metallicity and the star
formation history. We have slightly modified their relation including a
steepening of the synchrotron spectrum by $\Delta \alpha=0.5$ above a break
frequency of 20 GHz to take into account  electron ageing effects
\citep{BandayWolfendale1991}. The SFR-synchrotron luminosity relation then
writes:
\begin{equation}\label{eq:Lsync}
\bar{L}_{\rm sync}\simeq 1.9\times 10^{28}
\left(\frac{\hbox{SFR}}{\hbox{M}_{\odot}\hbox{yr}^{-1}}\right)
\left(\frac{\nu}{\hbox{GHz}}\right)^{-0.85}\left[1+\left({\nu\over 20\,\rm
GHz}\right)^{0.5}\right]^{-1}\, \hbox{erg}\,\hbox{s}^{-1}\,\hbox{Hz}^{-1}.
\end{equation}
The high-frequency synchrotron emission can be further suppressed at high $z$
because of the energy losses of relativistic electrons due to inverse Compton
scattering off the Cosmic Microwave Background (CMB) photons, whose energy
density increases as $(1+z)^{4}$ \citep{Norris2013, Carilli2008,
Murphy2009}. Also it should be noted that, at variance with free-free and UV
emissions which are a measure of the current SFR, the synchrotron emission is
delayed by $\sim 10^7\,$yr, the main sequence lifetime of massive stars whose
supernova explosions are responsible for the production of the relativistic
electrons \citep{Clemens2008}. At high $z$ the fraction of galaxies younger
than $\sim 10^7\,$yr, which are lacking relativistic electrons, hence the
synchrotron emission, increases.

Since these effects are not taken into account in our calculations because we
lack an adequate model for them, we may overestimate the faint counts and the
high-$z$ tails of the redshift distributions. The overestimate of the total
radio luminosity is however unlikely to be large because even at 1.4 GHz (and
more so at higher frequencies) at flux densities below the current
observational limits the free-free contribution  is comparable to the
synchrotron one (see Fig.~\ref{fig:1.4GHz_counts}).

In translating the SFR function into a synchrotron luminosity
function we need to take into account that the correlation between the SFR
and the radio luminosity has a substantial dispersion. Furthermore, it has
long been suggested that the synchrotron radiation from low SFR galaxies is
somewhat suppressed \citep{Klein1984,ChiWolfendale1990,PriceDuric1992},
although this view is controversial \citep{Condon1992}. A comparison of the
local SFR function with the local radio luminosity function at 1.4 GHz
\citep[dominated by synchrotron emission;][]{MauchSadler2007} shows that the
latter is flatter than the former at low luminosities. If the corresponding
portion of the SFR function was inferred from far-IR data one could argue
that it is overestimated since a fraction of the far-IR emission could be due
to dust heated by evolved stars. But in fact it was derived from H$\alpha$
measurements and therefore it is not liable to such a problem. To reconcile
the SFR function with the radio luminosity function we need to assume a
non-linear $L_{\rm sync}$--SFR relation. Following \citet{Massardi2010} we
adopt a relation of the form:

\begin{equation}\label{eq:Lsync_nonlin}
L_{\rm sync}(\nu)=\frac{L_{\star,\rm sync}(\nu)}{\left({L_{\star,\rm
sync}}/{\bar{L}_{\rm sync}}\right)^{\beta}+\left({L_{\star,\rm
sync}}/{\bar{L}_{\rm sync}}\right)},
\end{equation}

where $\bar{L}_{\rm sync}$ is given by eq.~(\ref{eq:Lsync}) and
$\beta = 3$. This equation converges to eq.~(\ref{eq:Lsync}) for
$\bar{L}_{\rm sync}\gg L_{\star,\rm sync} = 0.886 \bar{L}_{\rm sync}({\rm
SFR}=1\,M_\odot\,\hbox{yr}^{-1})$. At 1.4\,GHz, $L_{\star,\rm sync}\simeq
10^{28}\, \hbox{erg}\,\hbox{s}^{-1}\,\hbox{Hz}^{-1}$. Applying this relation
to the redshift dependent SFR functions presented in Sect.~\ref{sect:model}
(see Fig.~\ref{fig:SFRF}) and further adopting for the $L_{\rm radio}$--SFR
relation a dispersion of $\sigma_{\rm radio}=0.4\,$dex we get agreement with
the \citet{MauchSadler2007} luminosity function  (see
Fig.~\ref{fig:LLF1d4GHz}) as well as with the sub-mJy counts at 1.4\,GHz,
dominated by emission from star-forming galaxies.  The model counts, obtained
adding those of radio-loud AGNs yielded by the \citet{Massardi2010}
model\footnote{Tabulations of the radio-loud AGN counts yielded by the model
at several frequencies are publicly available at
http://w1.ira.inaf.it/rstools/srccnt/srccnt\_tables.html .}, are shown in
Fig.~\ref{fig:1.4GHz_counts}.

\subsection{Calibration of the relation between SFR and free-free emission}

We have rewritten the relation between SFR and free-free emission derived by
\citet{Murphy2012} as:
\begin{equation} \label{eq:Lff}
L_{\rm ff}=3.75\times 10^{26}
\left(\frac{\hbox{SFR}}{M_\odot/\hbox{yr}}\right) \,
\left(\frac{T}{10^4\,\hbox{K}}\right)^{0.3}\,
\hbox{g}(\nu,\hbox{T})\,\exp{\left(-{h\nu\over k\hbox{T}}\right)}
\end{equation}
where $T$ is the temperature of the emitting plasma and
$\hbox{g}(\nu,\hbox{T})$ is the Gaunt factor for which we adopt the
approximation of \citet{Draine2011}, which improves over earlier
approximations
\begin{equation}\label{eq:gaunt}
\hbox{g}(\nu,\hbox{T})=\ln\left\{\exp\left[5.960-\frac{\sqrt{3}}{\pi}\ln\left(Z_i{\nu\over
\hbox{GHz}} \left({T\over 10^4
\hbox{K}}\right)^{-1.5}\right)\right]+\exp(1)\right\},
\end{equation}
$Z_i$ being the charge of ions and $T$ their temperature. This equation
reproduces the \citet{Murphy2012} calibration at the calibration frequency
(33 GHz) for a pure hydrogen plasma ($Z_i=1$) and $T=10^4\,$K. At the other
frequencies relevant here the differences due to the different approximations
for the Gaunt factor are small. We will adopt the above values of $Z_i$ and
$T$ throughout the paper.

We have used eq.~(\ref{eq:Lff}) to convert the redshift-dependent SFR
functions to free-free luminosity functions. To compute the counts of
star-forming galaxies at 95\,GHz, where an observational estimate has been
obtained thanks to the SPT survey \citep{Mocanu2013}, we need to take into
account that the measured flux densities include, in addition to the
free-free emission, contributions from thermal dust and from the synchrotron
emission (see Fig.~\ref{fig:SEDsDusty}). All these contributions were taken
into account using the $L_{\rm ff}$--SFR and $L_{\rm sync}$--SFR relations
given above plus the $L_{\rm IR}$--SFR relation appropriate for each galaxy
population: eq.~(\ref{eq:LIR}) for late-type galaxies and
\begin{equation}\label{eq:SFR-IR}
L_{\rm IR} = 3.1 \times10^{43} \Big(\frac{\hbox{SFR}}{M_\odot\ {\rm
yr}^{-1}}\Big)\ {\rm erg\ s}^{-1},
\end{equation}
for proto-spheroidal galaxies \citep{Cai2013}. This calibration is
essentially identical to the empirical relation derived by
\citet{Murphy2012}.  To get down to any chosen frequency we used the SEDs for
the three populations of star-forming galaxies (``cold'' and ``warm''
late-type galaxies, and high-$z$ proto-spheroidal galaxies) considered by
\citet{Cai2013}, adding the free-free and synchrotron emissions given by
eqs.~(\ref{eq:Lff}) and (\ref{eq:Lsync}).

The model counts turned out to be substantially below those estimated by
\citet{Mocanu2013}. The statistical approach adopted by these authors,
however, did not appear completely reliable.  They used the
$\alpha^{150}_{220}$ spectral indices to classify their sources, taking
$\alpha^{150}_{220}=1.5$ as the boundary  between dust-dominated galaxies
($\alpha^{150}_{220}>1.5$) and radio-loud AGNs (synchrotron-dominated, in
their terminology). They computed the probability that their posterior value
for $\alpha^{150}_{220}$ was greater than the boundary value, and interpreted
it as the probability for a source to be dust-dominated. The 95\,GHz
differential counts were calculated as the sum of the probabilities
$P(\alpha^{150}_{220} > 1.5)$  that the sources in a given flux density
($S_{95\rm GHz}$) range are dusty. Since at 95\,GHz  the dusty galaxies are a
small fraction of detected sources, this approach may easily overestimate the
counts because, due to the large uncertainties on $\alpha^{150}_{220}$, many
radio loud AGNs have a non-zero probability of being dust-dominated. The sum
of these low probabilities may give a count comparable to, or even larger
than the count of the rare truly dusty galaxies.

This argument has motivated us to re-estimate the 95\,GHz counts of dusty
galaxies using a safer, although lengthier, approach. There are 406 95\,GHz
sources brighter than 12.6\,mJy, the 95\% completeness limit of the survey.
We have checked the SEDs of all of them, collecting the photometric data
available in the NASA/IPAC Extragalactic Database (NED) plus the Wide-field
Infrared Survey Explorer \citep[WISE;][]{Wright2010}
AllWISE\footnote{\url{http://wise2.ipac.caltech.edu/docs/release/allwise/}}
catalogue, the Planck Catalogue of Compact Sources
\citep{PlanckCollaborationXXVIII2013} and the IRAS catalogue. We found that
only 4  sources, all with $P(\alpha^{150}_{220} > 1.5)=1$, are indeed dusty
galaxies. They are listed in Table~\ref{tab:dusty_gal} and their SEDs are
shown in Fig.~\ref{fig:SEDsDusty}. A fifth source with $P(\alpha^{150}_{220}
> 1.5)\simeq 1$, SJ$043643-6204.4$, is only 0.368 arcmin away from the very
bright red giant Mira variable star R Doradus. Its SPT photometry may be
strongly contaminated by emission from dust shrouds that often form around
this kind of stars. For this reason we have discarded this source.

For comparison, the \citet{Mocanu2013} approach yields $\simeq 12$ dusty
galaxies with $S_{95\rm GHz}\ge 12.6\,$mJy. A substantial contribution to the
difference with our result comes from sources with $P(\alpha^{150}_{220} >
1.5)\ll 1$. The re-assessed integral counts are shown in
Fig.~\ref{conteggi95intnoconv}. It is worth mentioning that 3 out of the 4
confirmed dusty galaxies have low frequency radio luminosity in excess of
that expected from the mean relation between IR and 1.4 GHz luminosities. The
difference is about twice the dispersion of the relation between the
synchrotron luminosity and the SFR, yet within the range of values for
star-forming galaxies according to \citet{Yun2001}. It may thus be not
statistically significant, especially taking into account a possible
selection effect: the higher synchrotron luminosity helps bringing the
sources above the detection limit at 95 GHz. The parameter $q$ defined by
\begin{equation}
q=\log\left[{\hbox{FIR}/3.75\times 10^{12}\,\hbox{Hz}\over S_{1.4\rm
GHz}(\hbox{W}\,\hbox{m}^{-2}\,\hbox{Hz}^{-1})}\right],
\end{equation}
where
\begin{equation}
\hbox{FIR}(\hbox{W}\,\hbox{m}^{-2})=1.26\times 10^{-14}\left[2.58S_{60\mu{\rm
m}}(\hbox{Jy})+S_{100\mu{\rm m}}(\hbox{Jy})\right],
\end{equation}
has values ranging from 1.85 to 2.06 for the aforementioned galaxies,
somewhat smaller than the mean value of $q$ for star-forming galaxies
\citep[$q=2.34$ with a dispersion $\sigma=0.26$;][]{Yun2001}, yet larger than
the value $q_{\rm lim}=1.64$ proposed by \citet{Yun2001} as the boundary
between starburst-dominated and AGN-dominated radio emission. In any case, we
cannot rule out alternative explanations such as an additional contribution
from a weak active nucleus or excess radio emission arising from radio
continuum bridges and tidal tails not associated with star formation, similar
to what is observed for so-called ``taffy'' galaxies
\citep{Condon2002,Murphy2013}.

The SEDs of all the other sources are consistent with the 95\,GHz  flux
density being dominated by synchrotron emission and inconsistent with them
being dusty galaxies. As illustrated by Fig.~\ref{fig:SEDsRadio}, this is the
case also for the other 4 sources with $P(\alpha^{150}_{220} > 1.5)>0.6$,
including the one with $P(\alpha^{150}_{220} > 1.5)=0.91$. The radio-source
nature of all objects with lower values of $P(\alpha^{150}_{220} > 1.5)$ is
even clearer.

Figures~\ref{conteggi95intnoconv} and \ref{conteggi150_220} show that we
obtain good agreement with the new determination of SPT counts of dusty
galaxies at 95\,GHz as well as with those by \citet{Mocanu2013} at 150 and
220\,GHz. While the 95\,GHz counts are dominated by nearby galaxies, those at
the higher frequencies include a contribution from high-$z$ strongly lensed
proto-spheroidal galaxies; their contribution increases with decreasing flux
density (in the ranges covered by the SPT surveys) to the point of becoming
dominant at the SPT detection limits.

The model has been further tested at 4.8, 8.4, 15 and 30 GHz
(Fig.~\ref{fig:RadioCounts}). Not surprisingly, the free-free contribution
becomes increasingly important at higher and higher frequencies. While at 1.4
GHz the dominant emission is  synchrotron, at 8.4 GHz the free-free emission
takes over at sub-mJy levels.

\section{Predictions for surveys with the Square Kilometer Array (SKA) and its precursors}\label{sect:SKAcounts}

\subsection{Number counts and redshift distributions}

The plans for SKA Phase 1 (SKA1-MID) will likely include three
continuum surveys (in band 2), all at $\simeq 1$--1.4 GHz:  one over $1000\,
\hbox{deg}^2$ with rms $\simeq 1\,\mu$Jy/beam, a second one over
$30\,\hbox{deg}^2$ with rms $\sim 0.2\,\mu$Jy/beam, and a third over 1
deg$^2$ with rms $\sim 50$\,nJy/beam \citep{PrandoniSeymour2014}. In addition
there will possibly be a deep survey in band 5, at $\simeq 10\,$GHz, with rms
$\simeq 0.3\,\mu$Jy/beam over $1\,\hbox{deg}^2$, and an ultra-deep survey
again in band 5, at $\simeq 10\,$GHz with rms $\simeq  0.03\,\mu$Jy/beam over
$\simeq 30\,\hbox{arcmin}^2$ \citep[see also][]{Murphy2015}.

Deep surveys are also planned with the SKA precursors, like the Australian
SKA Pathfinder (ASKAP) and the South African MeerKAT \citep{Norris2013}. The
Evolutionary Map of the Universe (EMU) is a project that will use ASKAP to
make a deep ($5\,\sigma$ limit of $50\,\mu\hbox{Jy}\,\hbox{beam}^{-1}$) radio
continuum survey of the entire southern sky, extending as far north as
$+30^\circ$, i.e. covering about 30,000\,deg$^2$ at $\sim$1.4\,GHz. The
MeerKAT International GigaHertz Tiered Extragalactic Exploration (MIGHTEE)
Tier 2 will exploit the MeerKAT Phase 2 to observe over 35\,deg$^2$ down to a
$5\,\sigma$ limit of $5\,\mu\hbox{Jy}\,\hbox{beam}^{-1}$, again at $\sim
1.4\,$GHz.

The SKA1-MID, EMU and MIGHTEE $5\,\sigma$ flux density limits
for unresolved sources are indicated by vertical lines in
Fig.~\ref{fig:1.4GHz_counts}. The contributions of the various dusty galaxy
populations are shown by the different lines. The ``bump'' at tens of
$\mu$Jy's is due to  late-type galaxies at $z\simeq 1$--1.5. Proto-spheroidal
galaxies at higher $z$ become increasingly important at lower flux densities,
down to a few hundred nJy's.

The predicted redshift distributions for surveys at the flux density limits
mentioned above are shown in Figs.~\ref{fig:RedshiftDistr} and
\ref{fig:zdistband5}. The surface density of galaxies at $z\ge 6$ increases
rapidly with decreasing flux density. For the $5\,\sigma$ detection limits of
the 1.4\,GHz surveys mentioned above it is of $\simeq 10^{-2}$, 6.4, 148 and
$1013\,\hbox{deg}^{-2}$ for $S_{1.4\rm GHz}> 50$, 5, 1 and $0.25\,\mu$Jy
respectively, implying that the 3 SKA1-MID surveys are expected
to detect $\simeq 6,400$, 4,440 and 1013 $z\ge 6$ galaxies, respectively,
while the MIGHTEE and the EMU surveys should detect about 220 and 300
galaxies at these redshifts, respectively. Figure~\ref{fig:RedshiftDistr}
shows that the SKA1-MID surveys may reach even higher redshifts,
allowing us to get a glimpse of the cosmic SFR across the re-ionization
epoch, overcoming limitations by dust extinction. We expect the detection of
about 120, 163 and 58 galaxies at $z>8$  by the 3 SKA1-MID
surveys down to $S_{1.4\rm GHz} =5$, 1 and $0.25\,\mu$Jy respectively.

As illustrated by Fig.~\ref{fig:zdistband5}, the planned band 5 surveys will
also reach very high redshifts \citep[see also][]{Murphy2015}. We
predict the detection of $\simeq 16.5$ and $\simeq 5.5$  $z>6$ galaxies with
the deep and the ultra-deep survey, respectively. These estimates include the
contribution of dust emission that is significant for the highest redshift
galaxies (see Fig.~\ref{fig:SEDsDusty}). However, dust emission is likely
under-abundant in most galaxies  at $z> 6$
\citep{MunozLoeb2011,Cai2014,Capak2015} because there is probably not enough
time for a substantial dust enrichment of the galaxy interstellar medium.
However, galaxies with high SFRs and substantial dust extinction have
been reported \citep{Riechers2013,Watson2015}, and the timescale for dust
enrichment is poorly known \citep{Mancini2015}. SKA1-MID data
will provide key information also in this respect.

In Figure~\ref{fig:SKAcomp}  we compare the SKA1-MID potential in
measuring the evolution of the cosmic SFR to the results of \textit{Herschel}
and of the deepest UV and H$\alpha$ surveys. The planned SKA1-MID
surveys can detect galaxies with SFRs ranging from less than one $M_\odot/$yr
for $z\simlt 1$ to less than hundred $M_\odot/$yr for $z\simlt 7$. This is an
improvement by more than two orders of magnitude compared to
\textit{Herschel} results and will allow us to determine the dust-enshrouded
star formation history up to much higher redshifts. The deepest rest frame UV
surveys do better, but only for dust-free galaxies. The deepest ALMA
maps at 1.1 and 1.3\,mm \citep{Carniani2015} reach lower SFRs than SKA1-MID
at $z\simgt 3$, thanks to the negative K-correction. We note that ALMA is not designed to be a pure
survey instrument and has a very small field of view (the Full Width at Half
Maximum of the primary beam is $22''$ at 300 GHz); however it is interesting to compare its capabilities with the SKA (see Figure~\ref{fig:SKAcomp}).

\subsection{Strongly lensed galaxies}

Figures~\ref{fig:RedshiftDistr} and \ref{fig:zdistband5} also show that a
substantial fraction of the highest redshift galaxies that should be detected
by the foreseen surveys are strongly lensed (magnification $\mu \ge 2$)
protospheroids. The strongly lensed fraction at high $z$ increases with
increasing flux density limit, while the total counts rapidly decrease.
Adopting the SISSA profile \citep{Lapi2012} for the lens galaxies and an
upper limit $\mu_{\rm max}=30$ to the gravitational amplification, the model
yields 1195, 432, 101 and 7.6 strongly lensed galaxies per square
degree brighter than the SKA1-MID (band 2), the MIGHTEE and the
EMU $5\,\sigma$ detection  limits of 0.25, 1, 5 and 50\,$\mu$Jy respectively
(see Fig.~\ref{fig:negrello}). They are approximately the 0.5--0.6\% of the
total number of galaxies at these flux density limits (to be precise, the
strongly lensed fractions are 0.62\%, 0.52\%, 0.47\% and 0.63\%,
respectively).  The model with $\mu_{\rm max}=30$  was shown by
\citet[][their Fig.~2]{Bonato2014} to provide a good fit to the SPT counts of
dusty galaxies without an IRAS counterpart, interpreted as high-$z$ strongly
lensed galaxies. In the case of band 5 SKA1-MID surveys we
predict, over the survey areas, $\simeq 4$ and $\simeq 70$ strongly lensed
galaxies brighter than 0.15 and $1.5\,\mu$Jy respectively. They are about the
0.36\% of the total number of detected galaxies at both flux density limits.

To estimate the number of galaxies for which SKA1-MID can provide
direct evidence of strong lensing by detecting at least two images we have
turned to the classical Singular Isothermal Sphere (SIS) profile for lens
galaxies, since the application of the SISSA model for this purpose is much
more cumbersome. Details on the calculation are given in the
Appendix~\ref{appB}. We find that two images will be detected for about 30\%
of strongly lensed galaxies brighter than $0.25\,\mu$Jy at 1.4\,GHz; this
fraction decreases to about 20\% at a flux limit of $5\,\mu$Jy. This
means that high-resolution radio observations alone will allow us to directly
discern strong lensing for 20--30\% of the strongly lensed sources. Note that
the detection of multiple images can happen thanks to the high spatial
resolution of SKA1-MID surveys. It will not be generally possible with the
lower resolution of EMU/ASKAP and MIGHTEE/MeerKAT surveys.

\section{Synergies with optical/near-IR surveys.}\label{sect:synergies}

A full scientific exploitation of the data provided by the SKA surveys
requires redshift determinations of the detected galaxies, especially of the
most distant ones which carry information on the earliest stages of galaxy
formation. Several deep optical/near-IR imaging surveys covering substantial
sky areas are ongoing or foreseen.

In Fig.~\ref{fig:SKAobscured} we compare the minimum SFR of a dust-obscured
galaxy (whose SED is shown in the right-hand panel of Fig.~\ref{fig:SKA-JWST
SED}) detectable, as a function of redshift, by the deepest SKA1-MID survey with those detectable by the following surveys:

\begin{itemize}

\item The \textit{Euclid} deep survey covering an area of
    $40\,\hbox{deg}^2$ distributed over two fields close to the North and
    South ecliptic poles to $5\sigma$ magnitude limits for point source
    detection of 26\,mag in the Y, H and J bands and of 27.5\,mag in the
    VIS \citep{Laureijs2014}.

\item The Subaru Hyper Suprime-Cam (HSC) ultra-deep
    survey\footnote{www.subarutelescope.org/Projects/HSC/surveyplan.html}
    \citep{Miyazaki2012} covering two fields of $1.8\,\hbox{deg}^2$ each to
    $5\sigma$ point source AB magnitude limits of   $g= 28.1$, $r= 27.7$,
    $i= 27.4$, $z= 26.8$ and $y= 26.3$.

\item The 4 Large Synoptic Survey Telescope \citep[LSST;][]{Ivezic2008}
    deep drilling fields of $9.6\,\hbox{deg}^2$ each, aimed at reaching
    $5\sigma$ point source AB magnitude limits of 28\,mag in the $u$, $g$,
    $r$, $i$ and $z$ bands and of 26.8 in the $y$ band (see Sect. 9.8 of
    the LSST Science
    Book\footnote{www.lsst.org/files/docs/sciencebook/SH\_whole.pdf}).

\item The Deep-Wide Survey \citep[DWS;][]{Windhorst2009} to be carried out
    with NIRCam on the James Webb Space Telescope (JWST). The DWS will
    image the sky with up to 8 filters. A severe limitation to the
    exploitation of the SKA1-MID/DWS synergy is the smallness
    of the DWS area, 100\,arcmin$^2$, i.e. a factor of  $1/36\simeq 0.028$
    smaller than the area of the deepest SKA1-MID survey
    ($1\,\hbox{deg}^2$). We expect that the SKA1-MID will
    detect, within the DWS area, $\simeq 6,200$ galaxies at $z \leq 6$ but
    only $\simeq 14$ at $z \geq 6$.

\end{itemize}

\noindent The depth of the 4 LSST deep drilling fields is enough to
detect all SKA1-MID dust-obscured galaxies in at least 3 bands up
to $z\simeq 6$ and in at least one band up to $z\simeq 8$. Thus, photometric
redshift estimates can be obtained up to $z\simeq 6$. With the obvious
exception of the JWST/DWS the other surveys are somewhat less expedient. In
the case of the UV--bright SED (left-hand panel of Fig.~\ref{fig:SKA-JWST
SED}) all galaxies detected by the deepest SKA1{-MID} survey are easily
detected by the above optical/near-IR surveys in all bands, so that accurate
photometric redshift estimates can be obtained. Since, as argued above,
strong dust obscuration appears to affect only a minority of faint $z \geq 6$
galaxies, it may be expected that photometric redshift estimates will be
missing only for a small fraction of SKA1-MID high-$z$ galaxies.

One may then wonder what the SKA1-MID adds to the information
provided by deepest optical/near-IR surveys. Apart from the fact that the SFR
inferred from SKA data are immune to the effect of dust obscuration, even in
the case of no obscuration the SKA measurements are complementary to the
optical/near-IR ones. As noted in sub-sect.~\ref{sect:sync}, while the rest
frame UV emission measures the instantaneous SFR, the synchrotron emission
measures the SFR $\sim 10^7\,$yr earlier. Thus, in the absence of dust
obscuration, the comparison of the SFR inferred from (rest-frame) UV
photometry with that from low-frequency radio data, dominated by synchrotron,
yields a measure of the starburst age \citep{Clemens2008}. Starbursts younger
than $\sim 10^7\,$yr have a synchrotron emission decreasing with the
starburst age, so that the inferred SFR is lower and lower than that inferred
from UV data. By the same token, the comparison of SFRs derived from UV and
low-frequency radio measurements is informative on variations of the SFR on
timescales of $\sim 10^7\,$yr. A note of caution is in order however.
This comparison could not work at very high redshifts where the synchrotron
emission may be highly suppressed due to the strong inverse Compton losses
off the CMB \citep[see sub-sect.~\ref{sect:sync} and][]{Norris2013,
Carilli2008, Murphy2009}, so that even at relatively low frequencies the
radio emission may not be synchrotron dominated.

A rough redshift estimate for SKA1-MID galaxies missing better
information can be obtained from the expected correlation of the radio
spectral index with redshift. As $z$ increases the spectral index between the
frequencies $\nu_1$ and $\nu_2$ in the observer frame, $\alpha=
\log(S(\nu_1)/S(\nu_2))/\log(\nu_2/\nu_1)$, flattens with increasing $z$, due
to the increasing contribution of free-free compared to synchrotron. As an
example, Fig.~\ref{fig:alpha} shows the redshift dependence of the spectral
index between 1.4 and 4.8 GHz. The accurate photometric redshifts that can be
provided by the optical/near-IR surveys will allow us to better assess this
relation.

\section{Conclusions}\label{sect:conclusions}

We have produced detailed predictions for the counts and the redshift
distributions for planned surveys with the SKA1-MID and its
precursors, taking into account the contributions of the different
star-forming populations (late-type -- normal and starburst -- and
proto-spheroidal galaxies), allowing for the effect of gravitational lensing.
As mentioned in Sect.~\ref{sect:model}, the star formation is still
ongoing in late late-type galaxies, while it mostly occurred at $z\simgt 1.5$
in proto-spheroidal galaxies. Thus the dominant star forming population
gradually changes around  $z\sim 1.5$. The predictions rely on the redshift
evolution of the SFR function inferred from the wealth of data provided by
surveys at far-IR to mm wavevelengths as well as in the UV and in H$\alpha$.
These data sets complement each other by probing the dust obscured and
unobscured star formation, respectively.

In practice, we have exploited the models by \citet{Cai2013} for the dust
obscured star formation and by \citet{Cai2014} for the total SFR function,
including the unobscured one, at $z\simgt 2$. The missing contribution of the
unobscured star formation at $z\simlt 2$ has been added using a
phenomenological approach. The model extended in this way has been further
successfully tested against observational determinations at several redshifts
of the H$\alpha$ luminosity function, which is an independent measure of the
SFR function.

The SFR functions were converted into radio luminosity functions taking into
account both the synchrotron and the free-free emissions, using the
calibrations by \citet{Murphy2011,Murphy2012}. A comparison with the
\citet{MauchSadler2007} local luminosity function for star forming galaxies
at 1.4\,GHz showed evidence of a deviation from a linear relation between the
SFR and the synchrotron luminosity, in the sense that low-SFR galaxies are
under-luminous. This deviation was taken into account by means of a simple
analytic formula. Good agreement with the bright portion of the luminosity
function is achieved adopting a dispersion of $\sigma_{\rm radio}=0.4\,$dex
for the $L_{\rm radio}$--SFR relation.

This procedure, combined with the \citet{Massardi2010} model for
radio-loud AGNs, allowed us to reproduce the available source counts in the
range 1.4--30 GHz, but under-predicted the published SPT counts of dusty
galaxies at 95 GHz \citep{Mocanu2013}. However a careful  re-analysis of the
\citet{Mocanu2013} sample has shown that the surface density of dusty
galaxies was appreciably overestimated. The revised counts are in very good
agreement with model predictions, although the uncertainties are large, due
to the poor statistics. Our re-analysis also showed that the main
contribution to the observed 95 GHz flux density of local galaxies comes from
free-free emission, not from thermal dust as implied by some models.

With the radio luminosity function in hand, radio source counts and redshift
distributions at the relevant frequencies and detection limits were
straightforwardly computed. The highest redshift tails of the distributions
at the detection limits of planned SKA1-MID surveys were found to
include a substantial fraction of strongly lensed galaxies. We predict that
a survey down to $0.25\,\mu$Jy will detect about 1195 strongly lensed
galaxies per square degree, at redshifts of up to 10. For about 30\% of them
the SKA1-MID can detect 2 images, thus providing direct evidence
of strong lensing. The integral counts of strongly lensed galaxies are
roughly proportional to $S^{-0.75}$ in the flux density range
$0.01\,\mu\hbox{Jy}< S_{1.4\rm GHz}< 1\,\mu\hbox{Jy}$, implying that, at
fixed observing time, for the purpose of detecting these objects it is more
convenient to widen the covered area rather than going deeper. The counts
substantially steepen above $10\,\mu\hbox{Jy}$. Yet the EMU survey covering
30,000$\,\hbox{deg}^2$ to a $50\,\mu\hbox{Jy}$ $5\,\sigma$ detection limit is
expected to detect $\sim 2\times 10^4$ strongly lensed galaxies.

The redshift distributions also show that the SKA1-MID surveys,
and to some extent surveys with the SKA precursors, ASKAP and MeerKAT, will
allow us to probe the star formation history through the re-ionization epoch
without being affected by dust extinction. The deepest SKA1-MID
surveys, down to $0.25\,\mu\hbox{Jy}$, will extend the determination of the
SFR function by about two orders of magnitude compared to \textit{Herschel}
surveys up to the highest redshifts.

Finally, we have discussed the synergies between the deepest SKA1
-MID survey and deep ongoing or planned optical/near-IR surveys. We find
that the LSST deep drilling fields reach faint enough magnitudes to detect
essentially all SKA1-MID galaxies in at least 3 bands, allowing
reliable photometric redshift estimates. Exceptions may be heavily obscured
$z> 6$ galaxies that are however expected to be rare. The JWST/DWS can detect
also those in at least 6 filters, but its small area is a serious limitation.

\section*{Acknowledgements}
We are indebted to R. Aversa for having provided her analytic approximations
for the redshift-dependent SFR functions and acknowledge constructive
comments by an anonymous referee. This research has made use of the
NASA/IPAC Extragalactic Database (NED) which is operated by the Jet
Propulsion Laboratory, California Institute of Technology, under contract
with the National Aeronautics and Space Administration and of data products
from the Wide-field Infrared Survey Explorer, which is a joint project of the
University of California, Los Angeles, and the Jet Propulsion
Laboratory/California Institute of Technology, funded by the National
Aeronautics and Space Administration. We acknowledge financial support from
ASI/INAF Agreement 2014-024-R.0 for the {\it Planck} LFI activity of Phase
E2, from PRIN INAF 2012, project ``Looking into the dust-obscured phase of
galaxy formation through cosmic zoom lenses in the Herschel Astrophysical
Large Area Survey'' and from PRIN INAF 2014, project ``Probing the AGN/galaxy
co-evolution through ultra-deep and ultra-high resolution radio surveys''. Z.Y.C. is supported by China Postdoctral Science Foundation No. 2014M560515.
A.L. thanks SISSA for warm hospitality.

\begin{appendix}

\section{Evolution of the UV and H$\alpha$ luminosity functions of late-type galaxies}\label{appA}

For the UV ($\lambda = 1500\,$\AA) luminosity functions of late-type galaxies
we adopted the functional form chosen by \citet{Cai2013} for the IR
luminosity function of these galaxies:
\begin{equation}\label{eq:LF}
\Phi (\log L_{1500}, z) \hbox{d}\,\log L_{1500} = \Phi^*
\Big(\frac{L_{1500}}{L^*}\Big)^{1-\alpha}\times \exp \Big[- \frac{\log ^2
(1+L_{1500}/L^*)}{2\sigma^2} \Big] \hbox{d}\,\log L_{1500}.
\end{equation}
We obtained a sufficiently good description of the data using a simple pure
luminosity evolution model [$L^*(z) = L^*_0 (1+z)^{\alpha_L}$ up to $z = 1$,
see Fig.~\ref{fig:LFnu_1500A}]. The best-fit values of the parameters are
$\log[\Phi^*_0\,(\rm dex^{-1}\,Mpc^{-3})] = -2.150 \pm 0.095$,
$\log(L^*_0/L_\odot) = 9.436 \pm 0.119$, $\alpha = 1.477 \pm 0.050$, $\sigma
= 0.326 \pm 0.035$, and $\alpha_L = 2.025 \pm 0.063$. The free parameters are
constrained using the data at $z \leqslant 1$ by \citet{Wyder2005},
\citet{Treyer2005}, and \citet{Arnouts2005}. In Fig.~\ref{fig:LFnu_1500A} the
thin solid black lines show,  for comparison, the Schechter fits by
\citet{Arnouts2005}. The dot-dashed orange lines show the contributions of
proto-spheroidal galaxies derived adopting $A_{1500} = A_{1350} \cdot
k(1500\,\hbox{\AA})/k(1350\,\hbox{\AA}) = 0.936 A_{1350}$
\citep{Calzetti2000}.

As argued by \citet{Hopkins2001}, dust reddening of late-type galaxies
appears to be dependent on the SFR. Adopting their relation and the
\citet{Calzetti2000} extinction law, the extinction of the H$\alpha$ line is
given by:
\begin{equation}\label{eq:Ha_extinction}
A_{{\rm H}\alpha}(\log {\rm SFR}) = 6.536 \log \left[ \frac{0.797 \log(\rm
SFR) + 3.834}{2.88} \right].
\end{equation}
The corresponding UV extinction at 1500\,\AA, based on the
\citet{Calzetti2000} law is
\begin{equation}
A_{1500} = \frac{k(1500\,\angstrom)}{k(6563\,\angstrom)} \frac{E(B-V)_{\rm
star}}{E(B-V)_{\rm gas}} A_{\rm H\alpha} = 3.107 \times 0.44 \times A_{\rm
H\alpha} = 1.367 A_{\rm H\alpha},
\end{equation}\label{eq:UVext}
where we have taken into account that the color excess
appropriate for the stellar continuum, $E(B-V)_{\rm star}$, is related to
that for nebular gas emission lines, $E(B-V)_{\rm gas}$, by $E(B-V)_{\rm
star}=0.44 E(B-V)_{\rm gas}$ \citep{Calzetti2000}.

\begin{figure}
\includegraphics[width=\textwidth, height=10cm]{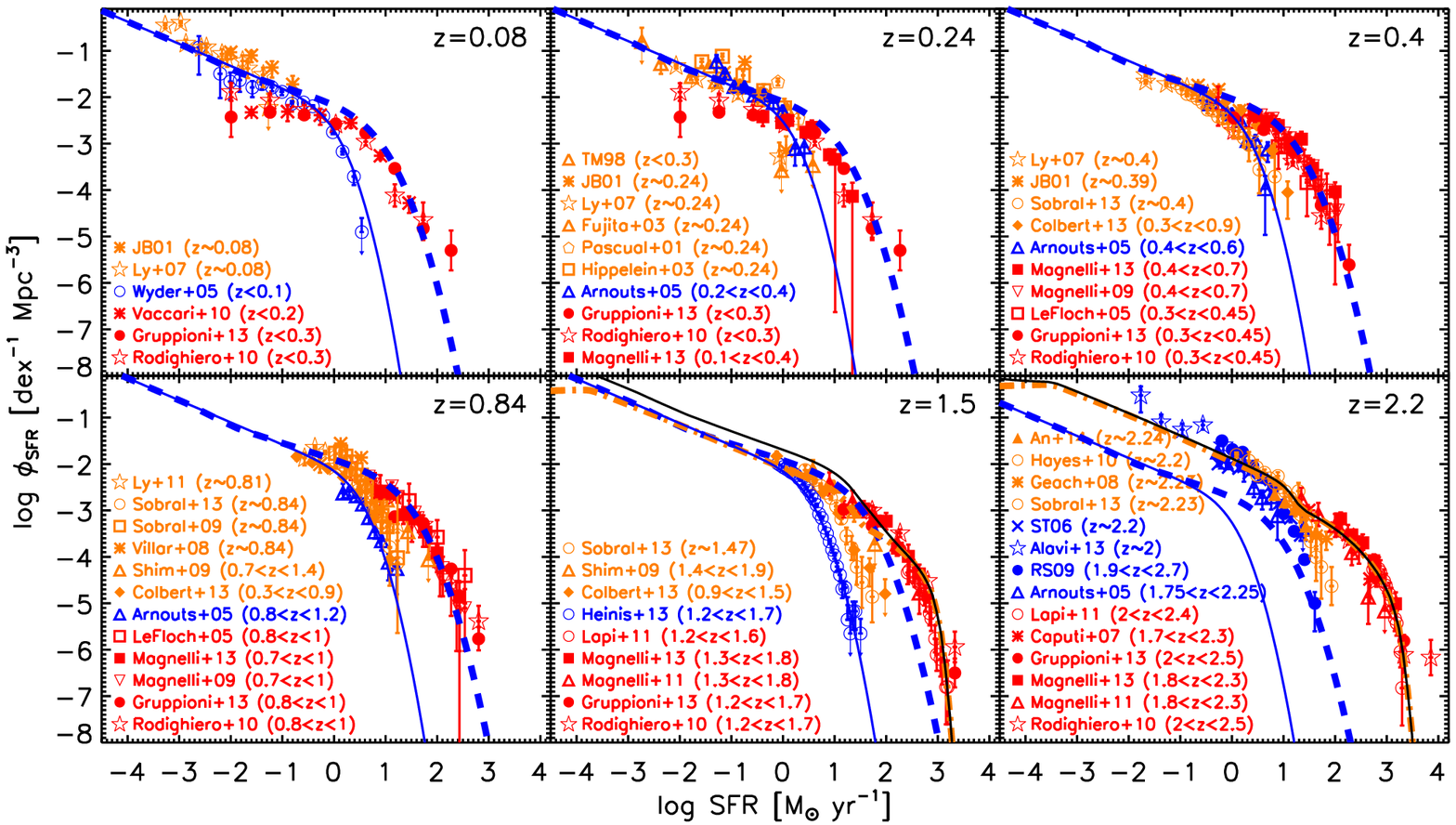}
\caption{SFR functions at several redshifts. The blue lines show the SFR
functions derived from the UV luminosity functions given by the
phenomenological model for late-type galaxies with and without the extinction
correction (dashed and thin solid lines, respectively). The dot-dashed orange
lines show the contributions given by the physical model for proto-spheroids
with a  minimum halo mass $M^{\rm min}_{\rm vir} = 10^{8.5}\,M_\odot$. The
thick black lines show the sum of the contributions from late-type galaxies,
after the extinction correction, and of proto-spheroids. The red data points
are estimates from far-IR measurements
\citep{LeFloch2005,Caputi2007,Rodighiero2010,Lapi2011,Gruppioni2013,Magnelli2011,Magnelli2013}.
The blue data points are from UV measurements
\citep{Wyder2005,Arnouts2005,Heinis2013,SawickiThompson2006,ReddySteidel2009,Alavi2014}.
The orange data points are from H$\alpha$ measurements
\citep{TresseMaddox1998,JonesBland-Hawthorn2001,Pascual2001,Fujita2003,Hippelein2003,Geach2008,Villar2008,Hayes2010,
Ly2007,Ly2011,Shim2009,Sobral2009,Colbert2013,Sobral2013,An2014}. 	 The conversions of attenuated UV and
H$\alpha$ luminosities to SFRs are done using the \citet{KennicuttEvans2012}
calibrations, i.e., $\log ({\rm SFR}/[M_\odot\,{\rm yr}^{-1}]) = \log (L_{\rm
FUV}/[L_\odot]) - 9.825$ and $\log ({\rm SFR}/[M_\odot\,{\rm yr}^{-1}]) =
\log (L_{\rm H\alpha}/[{\rm erg\,s^{-1}}]) - 41.270$, respectively. The
conversion of the IR luminosity to SFR was done using eq.~(\ref{eq:LIR})
which takes into account that a substantial contribution to dust heating in
galaxies with low SFRs comes from old stellar populations. Comparing the SFRs
estimated from the UV or from the H$\alpha$ luminosities with those inferred
from the IR luminosities it is apparent that the former are increasingly
underestimated with increasing SFR and redshift, consistent with the findings
by \citet{Reddy2012} and \citet{Burgarella2013}.} \label{fig:SFRF}
\end{figure}

\begin{figure}
\centering
\includegraphics[width=0.8\textwidth]{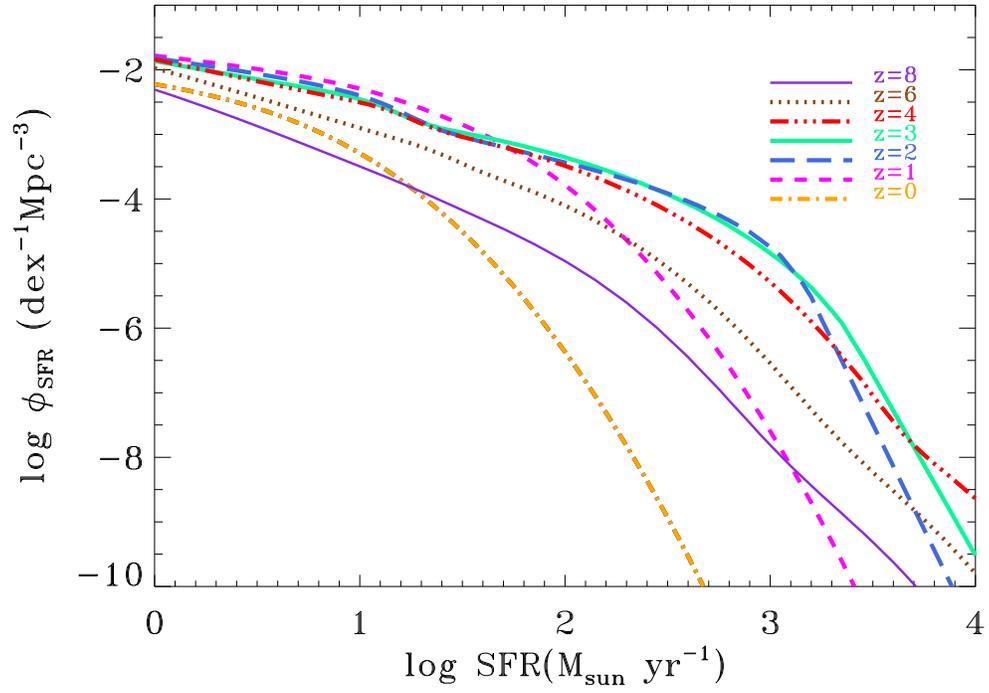}
\caption{Synoptic view of the evolution of the cosmic SFR function as given
by the model described in Section~\ref{sect:model}. The luminosity function at each
redshift includes the contributions of normal late-type, starburst and
proto-spheroidal galaxies.}\label{fig:SFRevol}
\end{figure}

\begin{figure}
\centering
\includegraphics[width=\textwidth, angle=0]{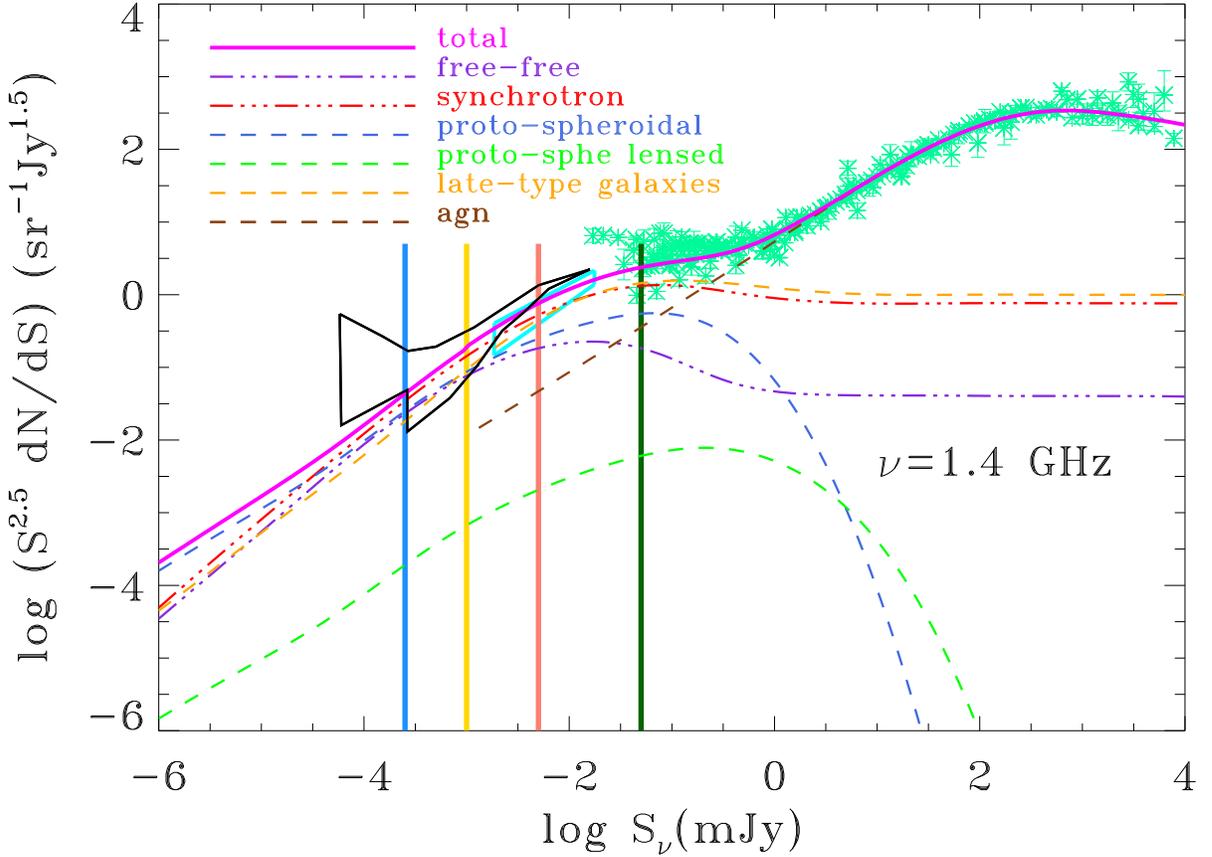}
\caption{\small Euclidean normalized differential counts at 1.4 GHz compared
with model predictions. Above $\sim 1\,$mJy the counts are dominated by
canonical radio-loud active galactic nuclei (AGNs) while star-forming
galaxies take over at fainter flux densities. The dashed brown line shows the
model for radio AGNs by \citet{Massardi2010}. The other dashed lines show the
contributions of the star-forming populations considered in this paper
(unlensed and strongly lensed proto-spheroids and late-type galaxies). The
two triple-dot dashed lines show the total synchrotron and free-free
emissions from these populations. References for the data points can be found
in \citet{DeZotti2010}. The cyan and the black polygons show the
ranges of 1.4 GHz counts consistent with the $P(D)$ distributions by
\citet{Condon2012} and \citet{Vernstrom2014}, respectively. Note that these
$P(D)$ analyses rule out the sub-mJy counts as high as the
\citet{OwenMorrison2008} estimates. As argued by \citet{Condon2012} the high
counts may result from the large and difficult corrections for the effects of
partial resolution; the quoted $P(D)$ data do not need significant resolution
corrections.  The vertical lines correspond, from left to right, to the
$5\,\sigma$ detection limits of the planned SKA1-MID surveys to
unresolved sources: $0.25\,\mu$Jy (blue),  $1\,\mu$Jy (yellow), $5\,\mu$Jy
(pink). Given the resolution of these surveys \citep[$\simeq
0.5''$;][]{PrandoniSeymour2014} it is likely that a fraction of sources
(especially bright, nearby galaxies) will be resolved out so that the surveys
will not be 100\% complete at the quoted limits. However, the array
configuration proposed, which delivers an essentially flat point source
sensitivity over a large range of synthesized beams, will overcome this
problem. The $5\,\mu$Jy limit applies also to the Tier~2 MIGHTEE survey
(MeerKAT), while the dark green line shows the EMU (ASKAP) survey $5\,\sigma$
limit. Again the limits  refer to unresolved sources. The MIGHTEE and EMU
surveys have a lower resolution than the SKA1-MID ones: $8.3''$--$3.5''$ in
the case of Tier~2 MIGHTEE
(http://www.ast.uct.ac.za/arniston2011/vdheyden.pdf) and $10''$ in the case
of EMU \citep{Norris2011}.} \label{fig:1.4GHz_counts}
\end{figure}

\begin{figure}
\centering
\includegraphics[width=0.8\textwidth]{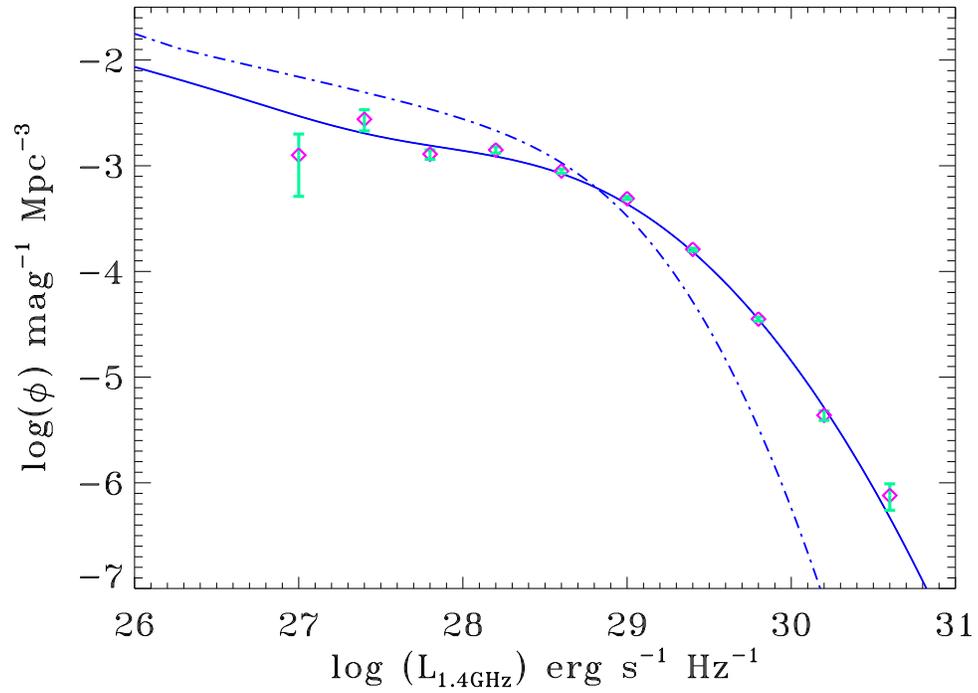}
\caption{Synchrotron luminosity function of local star-forming
galaxies derived from the local SFR function using the non-linear $L_{\rm
radio}$--SFR relation  (cf. eq.~\ref{eq:Lsync_nonlin}) and a dispersion of
$\sigma_{\rm radio}=0.4\,$dex around the mean relation, compared with the
observational determination by \citet{MauchSadler2007}. The luminosity
function obtained assuming a simple proportionality between $L_{\rm radio}$
and the SFR [eq.~(\ref{eq:Lsync})] is also shown for comparison (dot-dashed
line). The dispersion $\sigma_{\rm radio}$ raises the bright portion of the
luminosity function while eq.~(\ref{eq:Lsync_nonlin}) flattens the faint
part.}\label{fig:LLF1d4GHz}
\end{figure}

\begin{figure}
\hspace{+0.0cm}
{\includegraphics[width=0.48\textwidth, angle=0]{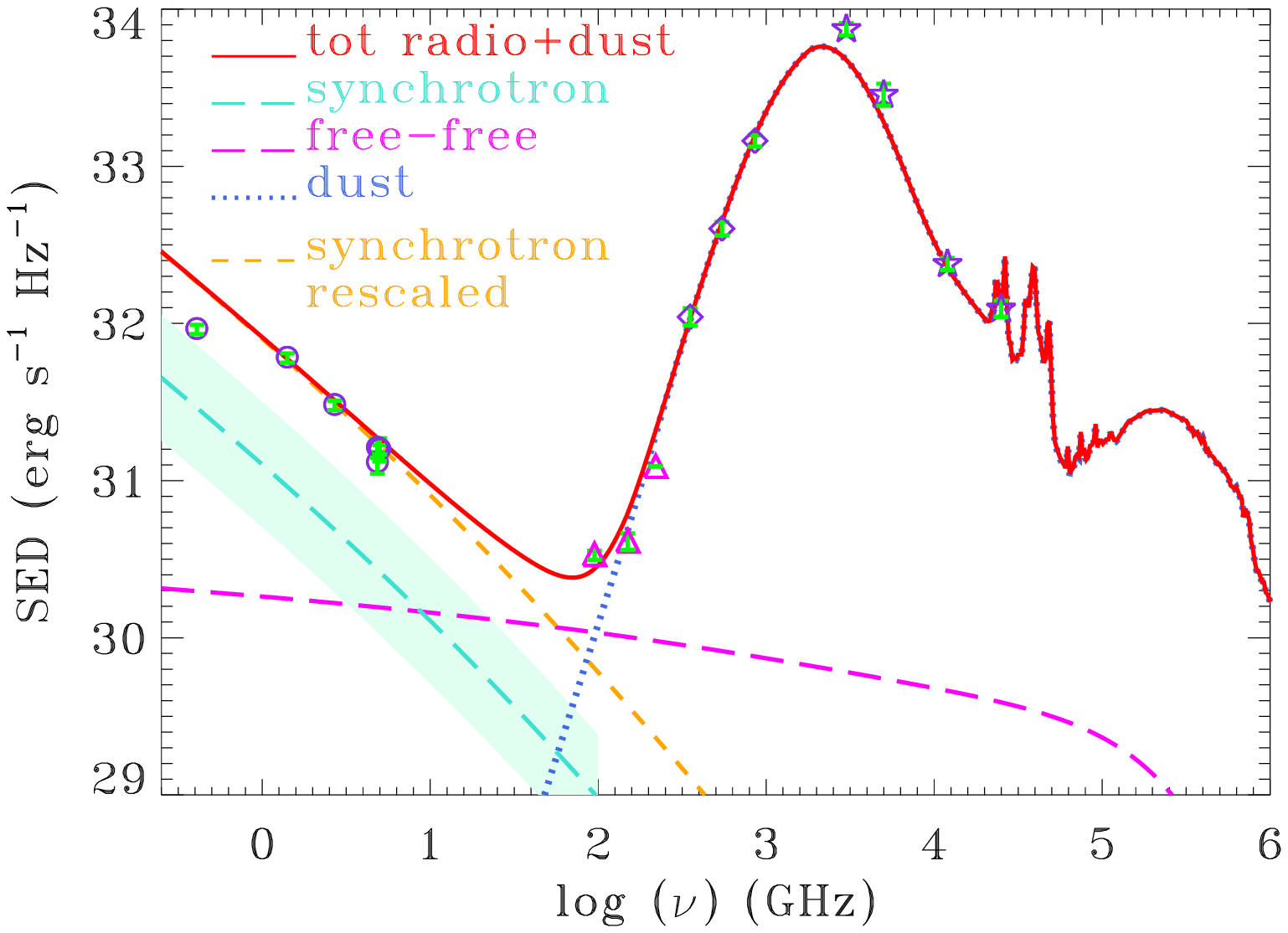}
\includegraphics[width=0.48\textwidth, angle=0]{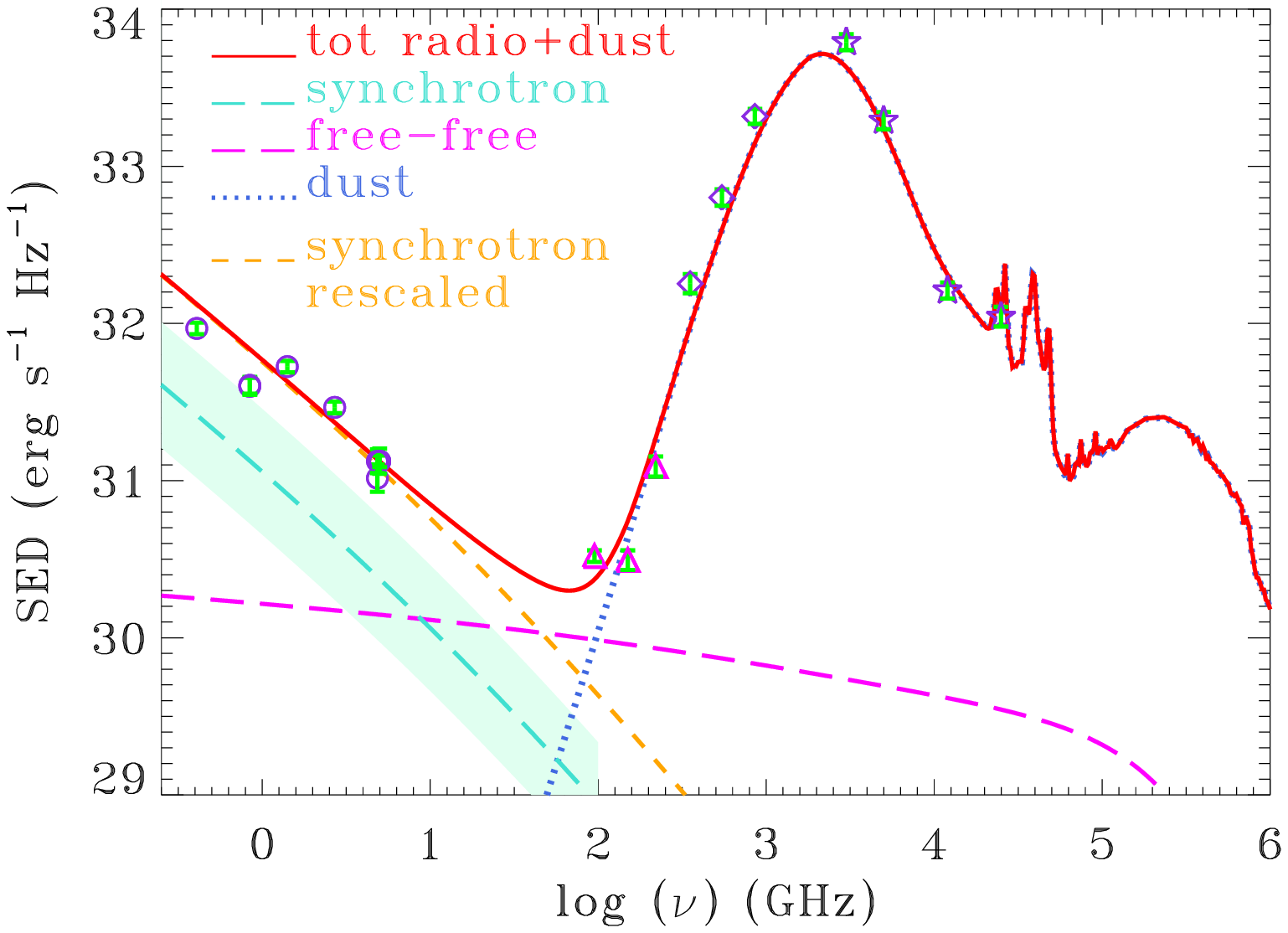}
\includegraphics[width=0.48\textwidth, angle=0]{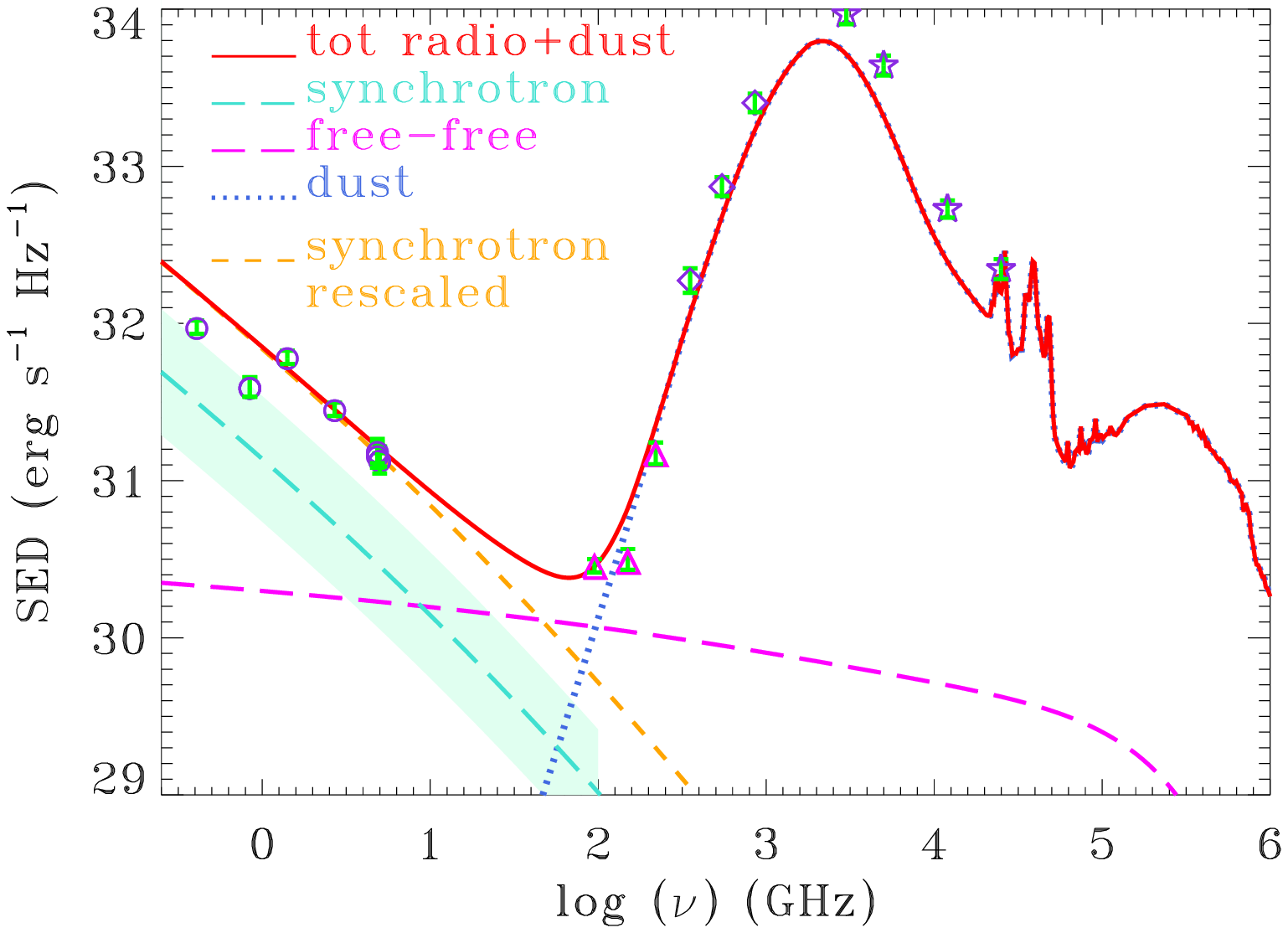}
\includegraphics[width=0.48\textwidth, angle=0]{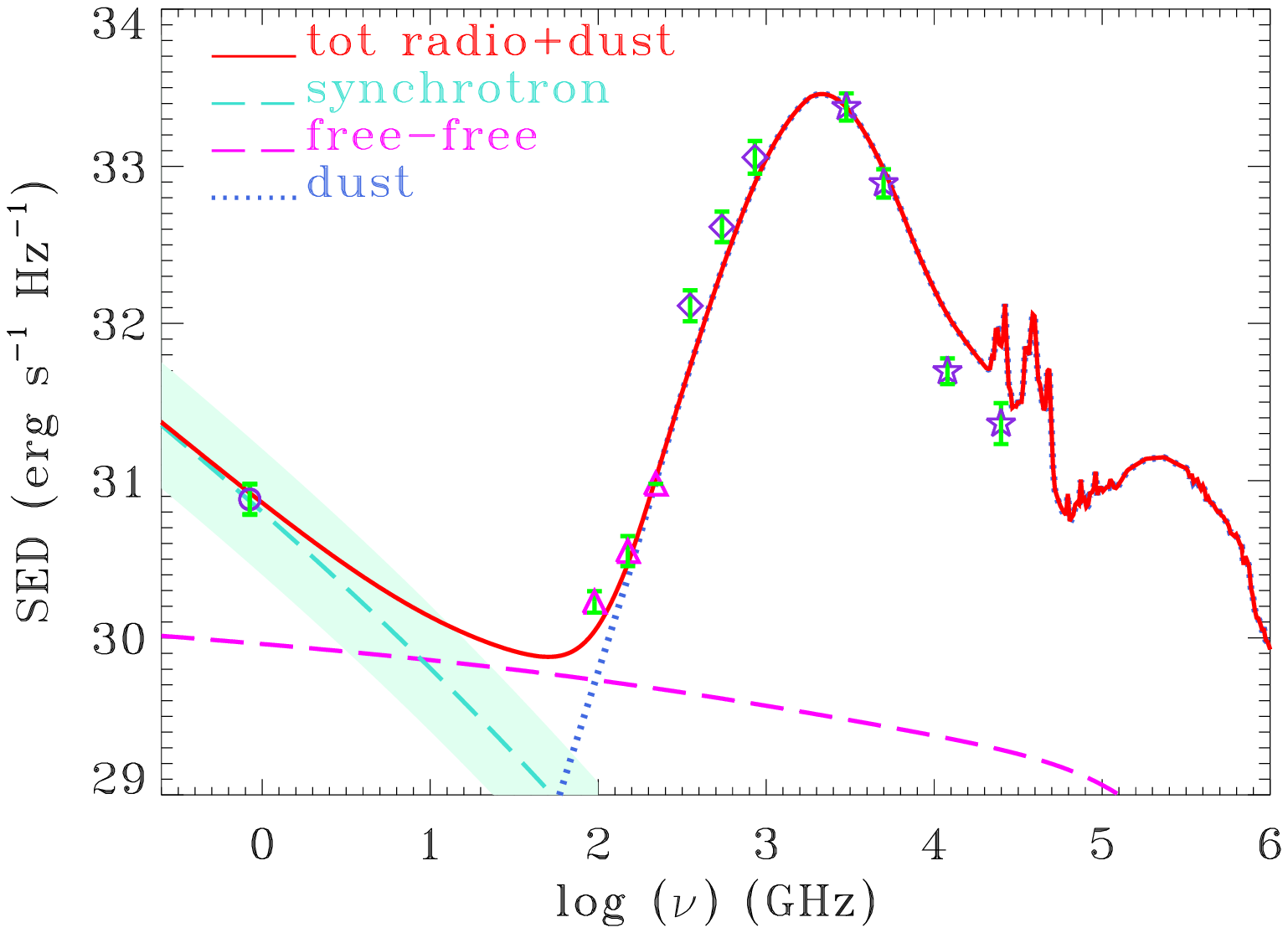}}
\caption{SEDs of 95\,GHz selected dusty galaxies in the SPT sample by
\citet{Mocanu2013}:  SPT-SJ041736-6246-8 (top left), SPT-SJ041959-5456-2 (top right), SPT-SJ044540-5914-7 (bottom left), SPT-SJ213629-5433-4 (bottom right). The red curve is a fit made by eye summing the
\citet{Cai2013} SED  of a cold late-type galaxy (dotted blue line), chosen
because the 4 galaxies are all normal disc galaxies, with the free-free
emission given by eq.~(\ref{eq:Lff}) for the same SFR (long-dashed magenta
line) and with the synchrotron emission matching the low frequency data. For
three of the galaxies, such synchrotron emission (short-dashed orange line)
is higher than that given by eq.~(\ref{eq:Lsync}), represented by the
long-dashed green line. The difference is about twice the dispersion
of the relation between the synchrotron luminosity and the SFR (shaded green
band; see sub-sect.~\ref{sect:sync}), yet within the range of values for
star-forming galaxies according to \citet{Yun2001}. It may thus be not
statistically significant, especially taking into account a possible
selection effect: the higher synchrotron luminosity helps bring the
sources above the detection limit at 95 GHz. However, we cannot rule out
alternative explanations such as an additional contribution from a weak
active nucleus or excess radio emission arising from radio continuum bridges
and tidal tails not associated with star formation, similar to what is
observed for so-called ``taffy'' galaxies \citep{Condon2002,Murphy2013}.
Data points: low frequency radio flux densities from the NED (open circles); SPT
flux densities (triangles); \textit{Planck} flux densities  (open diamonds);
IRAS flux densities (open stars). }
 \label{fig:SEDsDusty}
\end{figure}

\begin{figure}
\centering
\includegraphics[width=0.9\textwidth, angle=0]{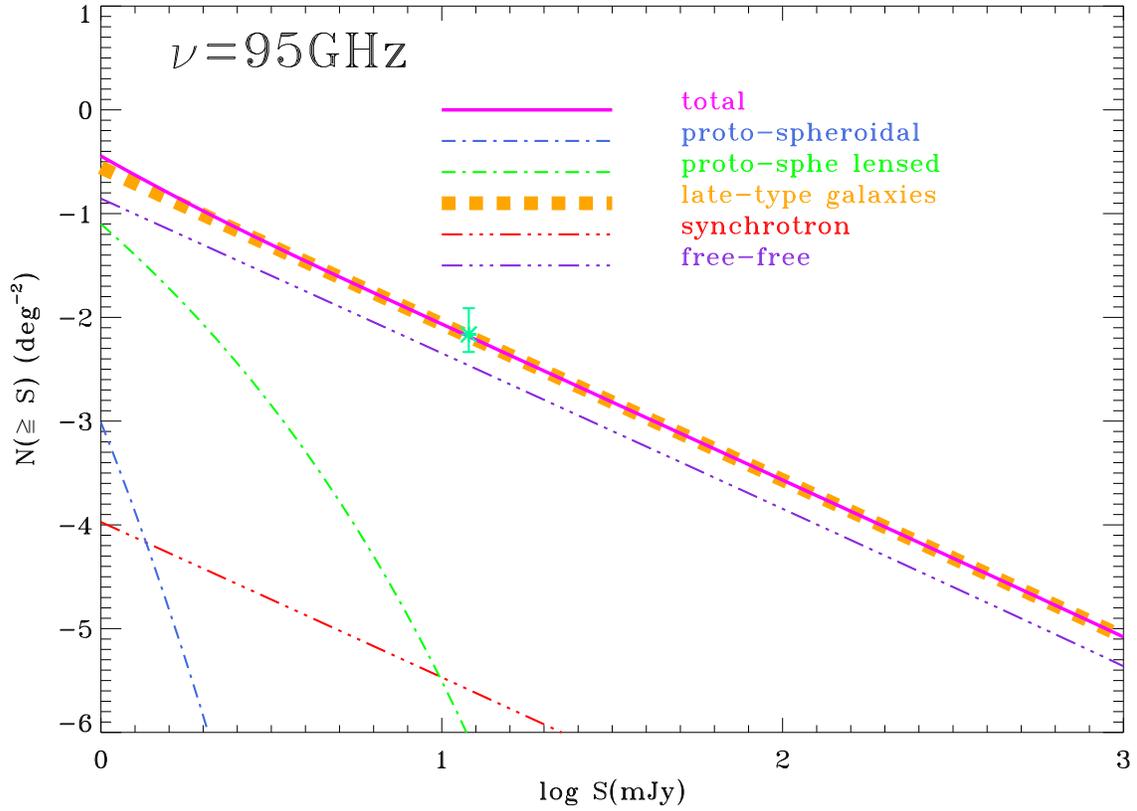}
\caption{Our re-assessment of the integral counts of dusty galaxies at
$\nu$=95 GHz (asterisks with error bars) compared with the expectations from
the \citet{Cai2013} model, using for each source populations the complete
SEDs that include dust, synchrotron and free-free emissions. The various
lines correspond to the contributions from the different source populations
and from the different emission processes, identified in the legend inside
the figure. }\label{conteggi95intnoconv}
\end{figure}

\begin{figure}
\hspace{+0.0cm}
\includegraphics[width=0.48\textwidth, angle=0]{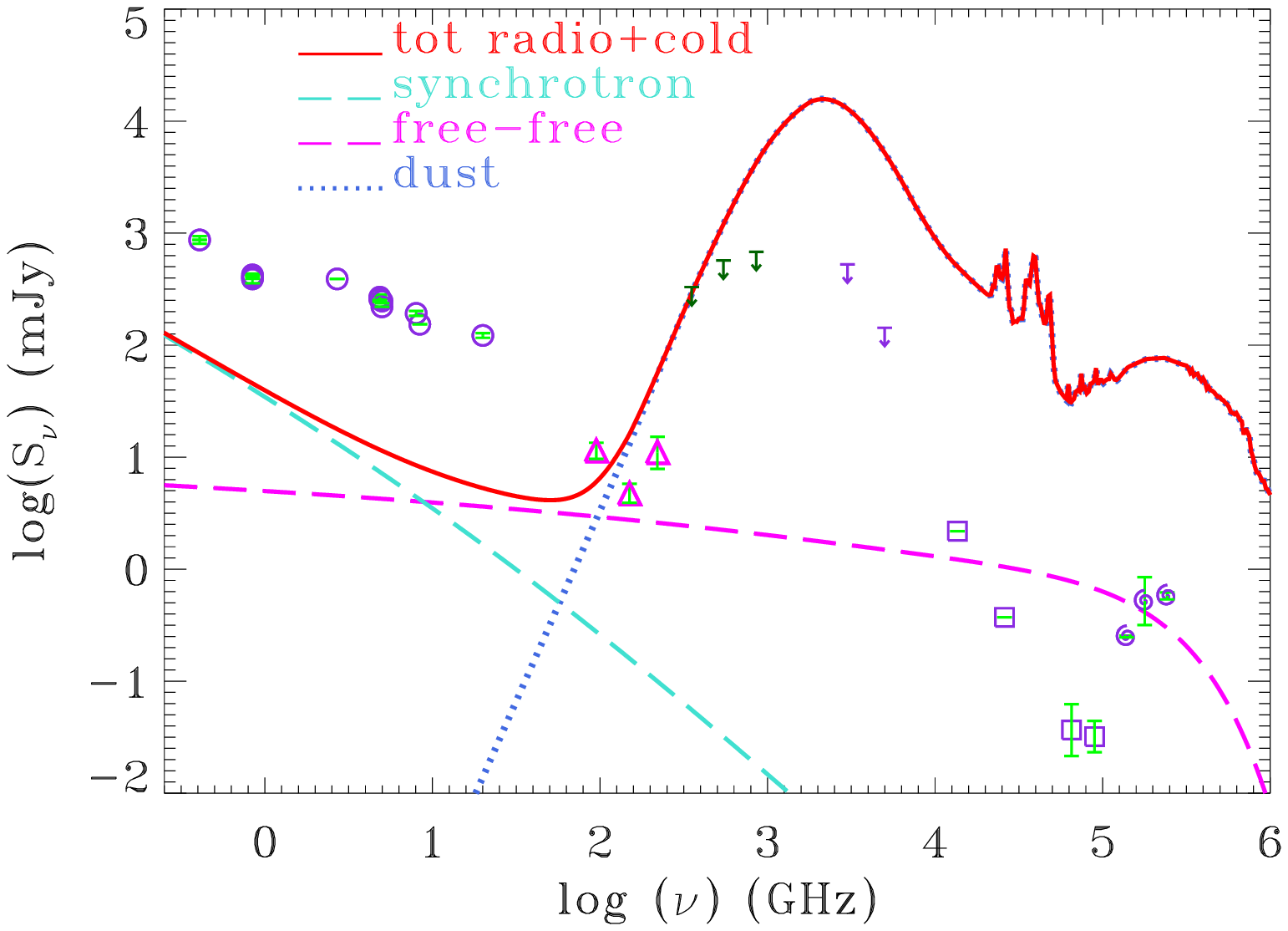}
\includegraphics[width=0.48\textwidth, angle=0]{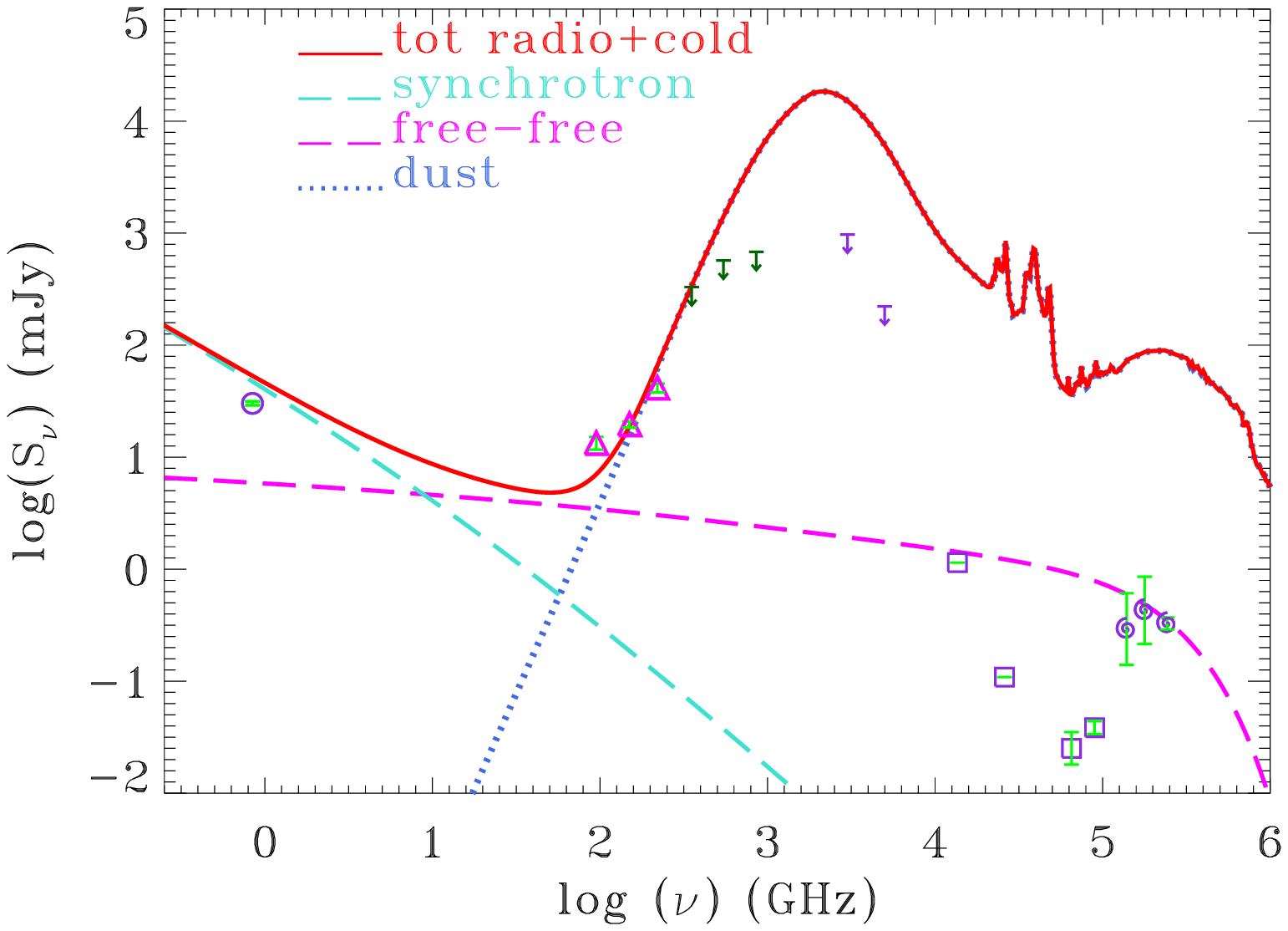}
\includegraphics[width=0.48\textwidth, angle=0]{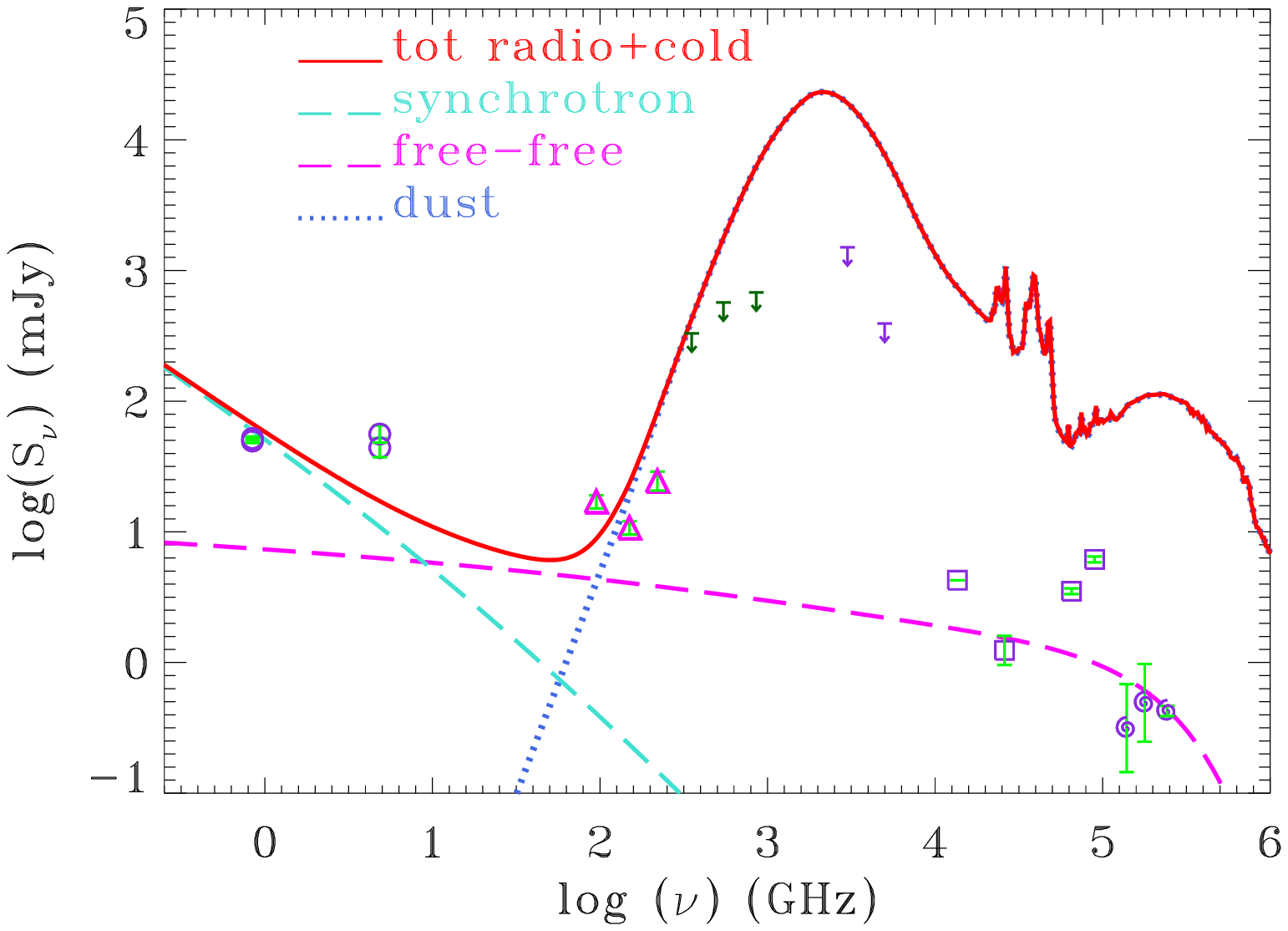}
\includegraphics[width=0.48\textwidth, angle=0]{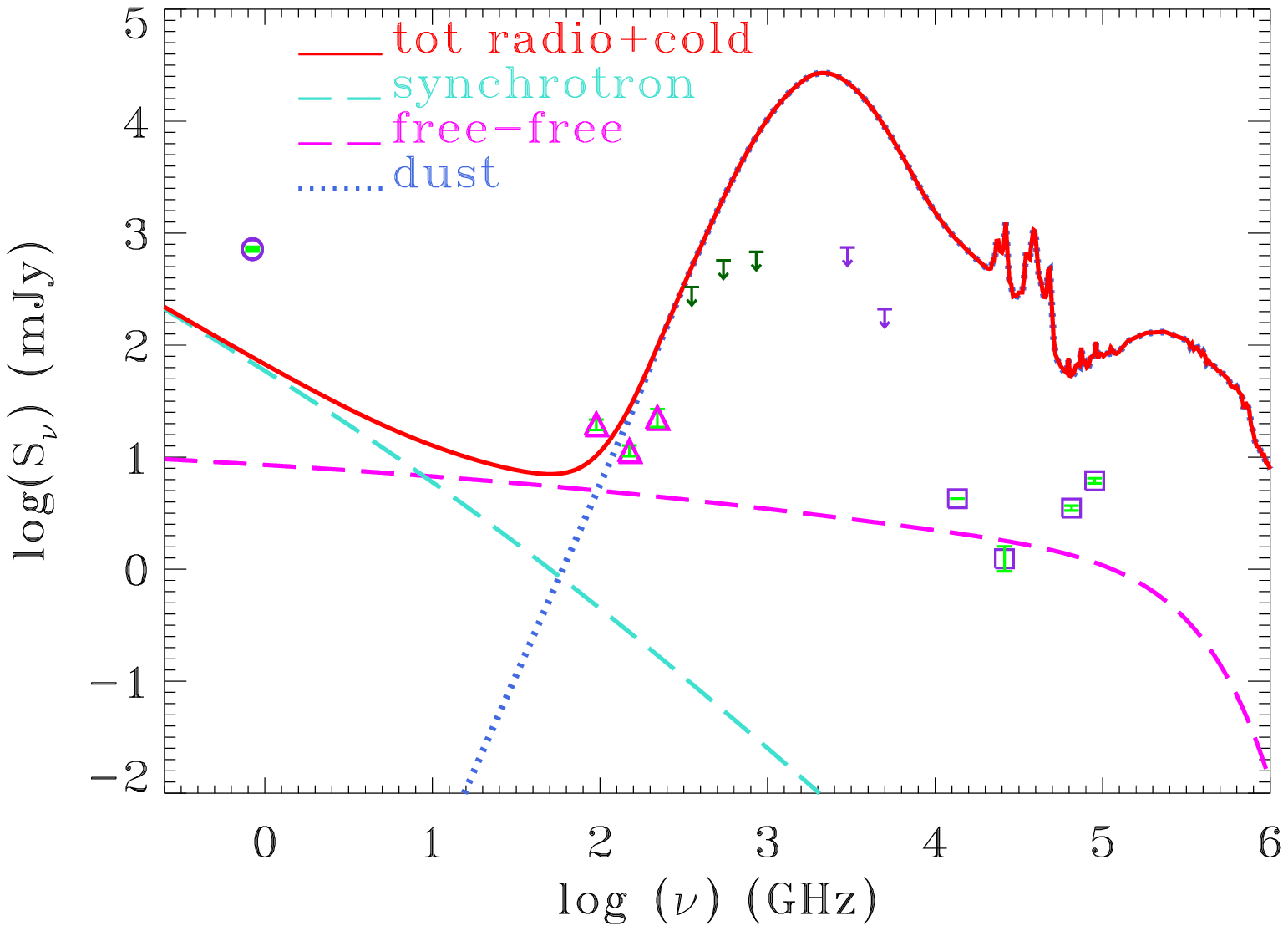}
\caption{SPT sources selected at 95\,GHz that have $1> P(\alpha^{150}_{220} >
1.5)>0.6$: SPT-SJ043651-5841-1 (top left), SPT-SJ203730-6513-3 (top right), SPT-SJ204101-5451-4 (bottom left), SPT-SJ213406-5334-3 (bottom right). Their SEDs are incompatible with those of star-forming galaxies,
illustrated by the solid red line, and consistent with being radio sources.
Data points: low frequency radio flux densities from the NED (open circles); SPT
flux densities (triangles); WISE flux densities (open squares); 2MASS flux
densities (spirals); in violet and dark green are IRAS and
\textit{Planck} upper limits, respectively. We have adopted as
\textit{Planck} upper limits the 90\% completeness limits of the
\textit{Planck} Catalogue of Compact Sources
\citep{PlanckCollaborationXXVIII2013} in the ``extragalactic zone''.}
 \label{fig:SEDsRadio}
\end{figure}

\begin{figure*}
\includegraphics[width=0.48\textwidth, angle=0]{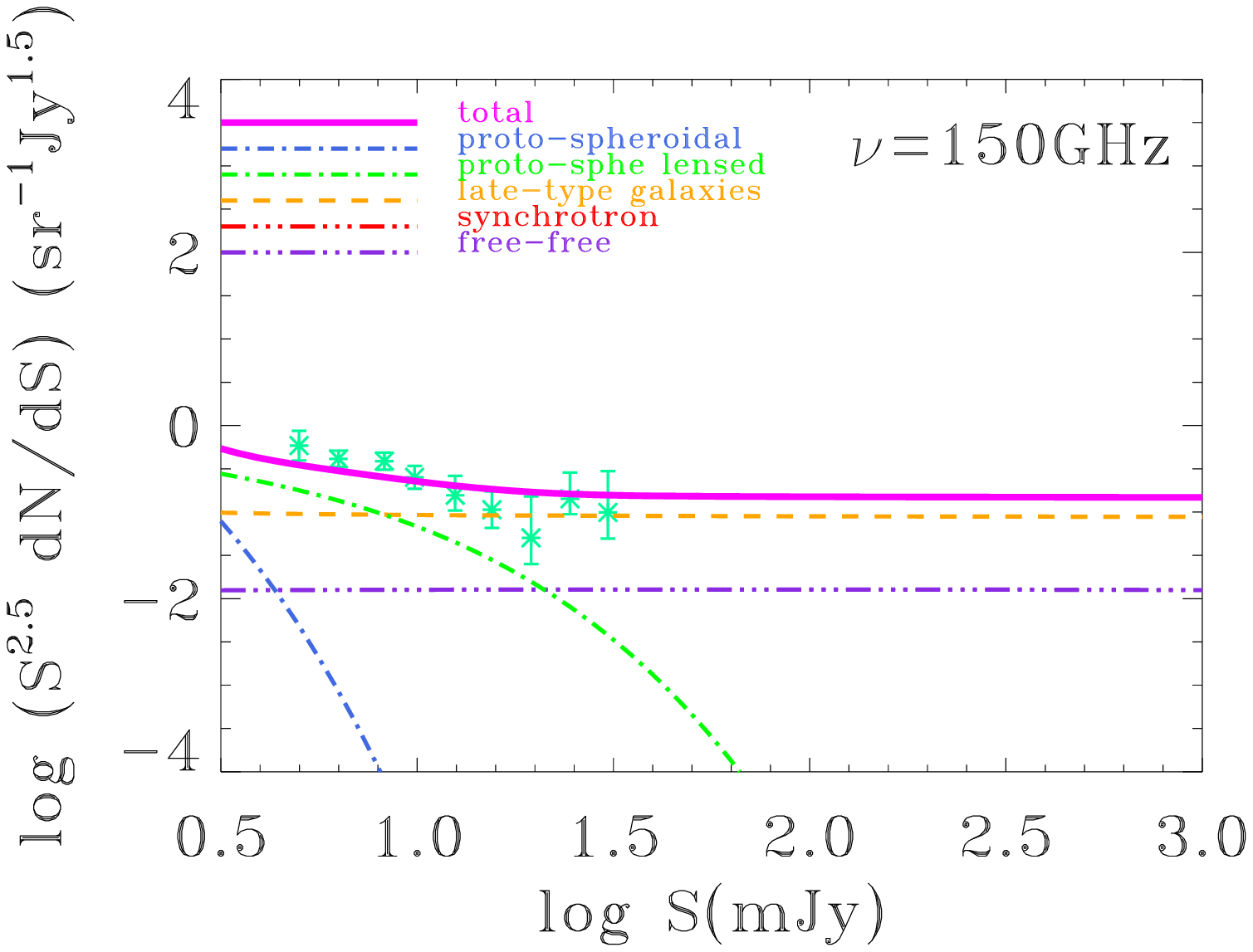}
\includegraphics[width=0.48\textwidth, angle=0]{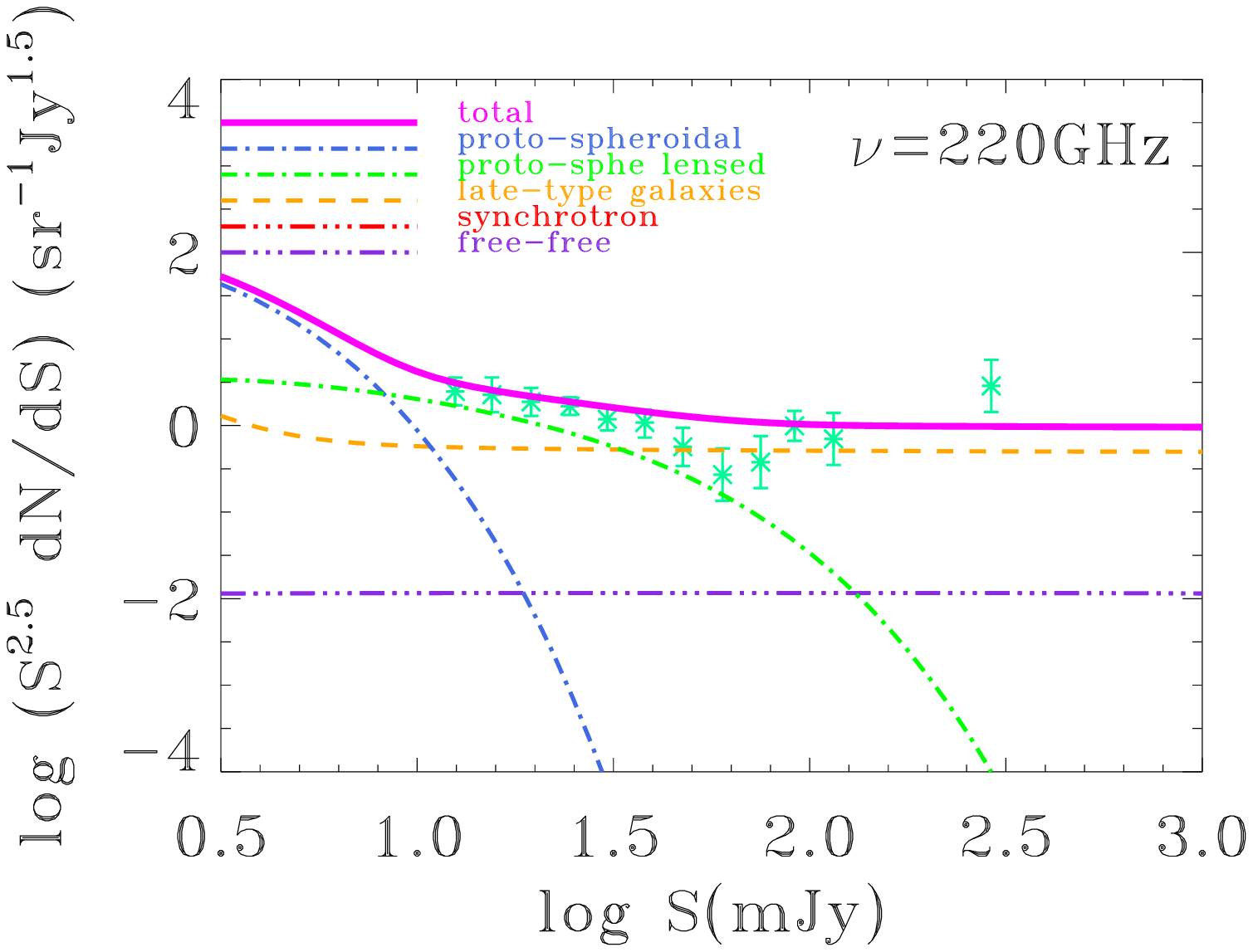}
\caption{Euclidean normalized differential counts of dusty galaxies at 150
and at $220\,$GHz (left and right panel, respectively). The asterisks with
error bars are the observational determinations by \citet{Mocanu2013}. The
various lines correspond to the contributions from the different emission
processes and from the different galaxy populations, identified in the legend
inside the panels. The lines corresponding to the synchrotron
emission are below the lower limit of the y-axis. Note that in the considered
flux density ranges the counts of star-forming galaxies seen via their radio
emission are made by nearby sources and have thus an Euclidean slope. On the
contrary, counts due to unlensed and lensed proto-spheroidal galaxies, seen
via their dust emission, are due to strongly evolving high-$z$ objects.}
\label{conteggi150_220}
\end{figure*}

\begin{figure*}
\centering
\hspace{+0.0cm}
\includegraphics[width=1\textwidth, angle=0]{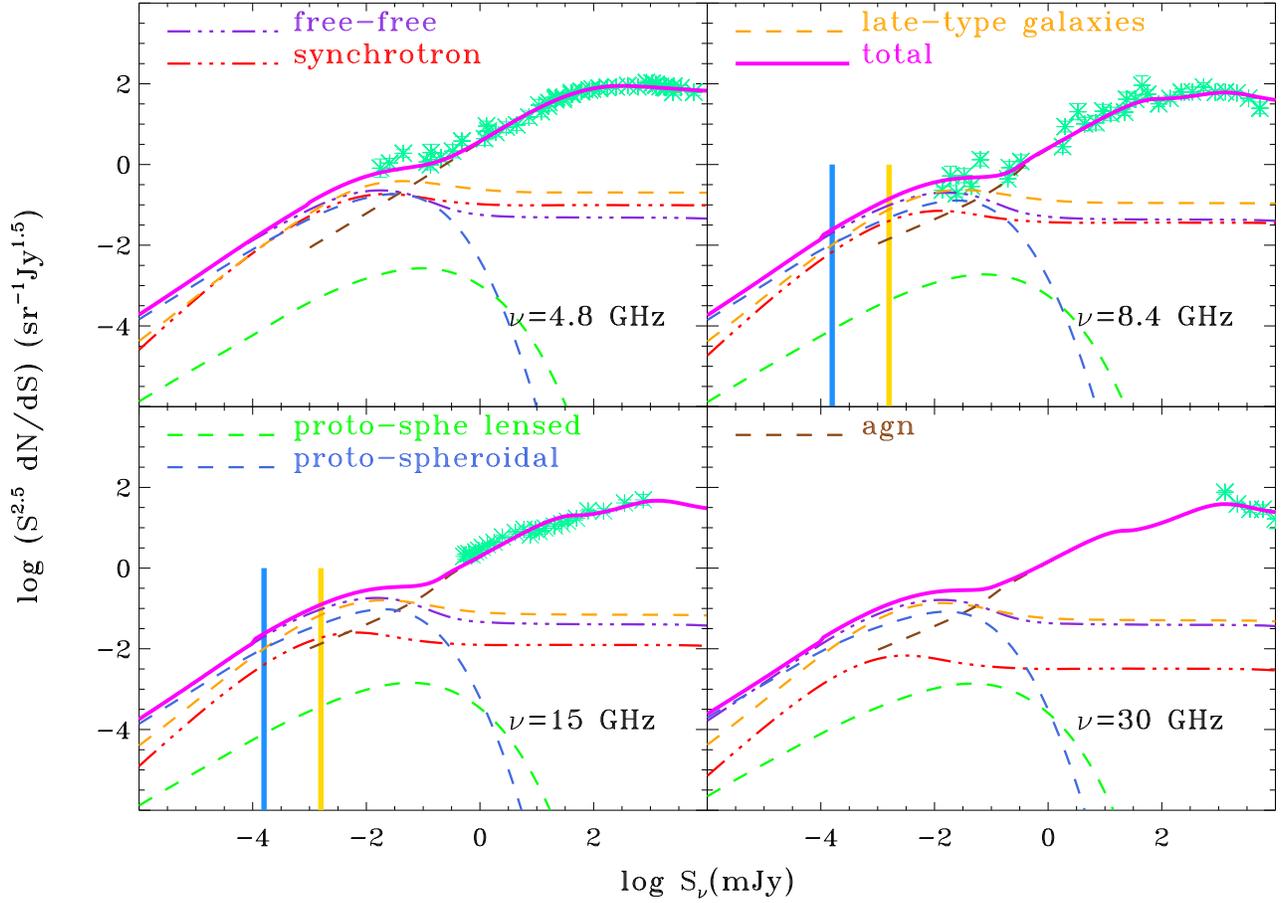}
\caption{Predicted versus observed counts at 4.8, 8.4 and 30 GHz. Dusty
galaxies come up at sub-mJy flux density levels and their counts are
accounted for by the model. At higher flux densities the counts are dominated
by canonical, AGN powered radio sources. The models shown are from
\citet{Massardi2010} at 1.4, and 4.8 GHz and from \citet{DeZotti2005} at 8.4
and 30 GHz. The various lines correspond to the contributions from the
different source populations (dashed yellow lines: late-type galaxies; dashed
blue lines: unlensed proto-spheroidal galaxies; dashed green lines: strongly
lensed proto-spheroidal galaxies; dashed brown lines: radio-loud AGNs; solid
magenta line: total) and from the different emission processes (triple-dot
dashed lines, red for synchrotron, violet for free-free). The solid vertical
lines indicate the predicted flux density limits, for unresolved
sources, of possible deep (yellow) and ultra-deep (blue) band 5 surveys with
SKA1-MID.}
 \label{fig:RadioCounts}
\end{figure*}

\begin{figure}
\centering
\hspace{+0.0cm}
\includegraphics[width=1\textwidth, angle=0]{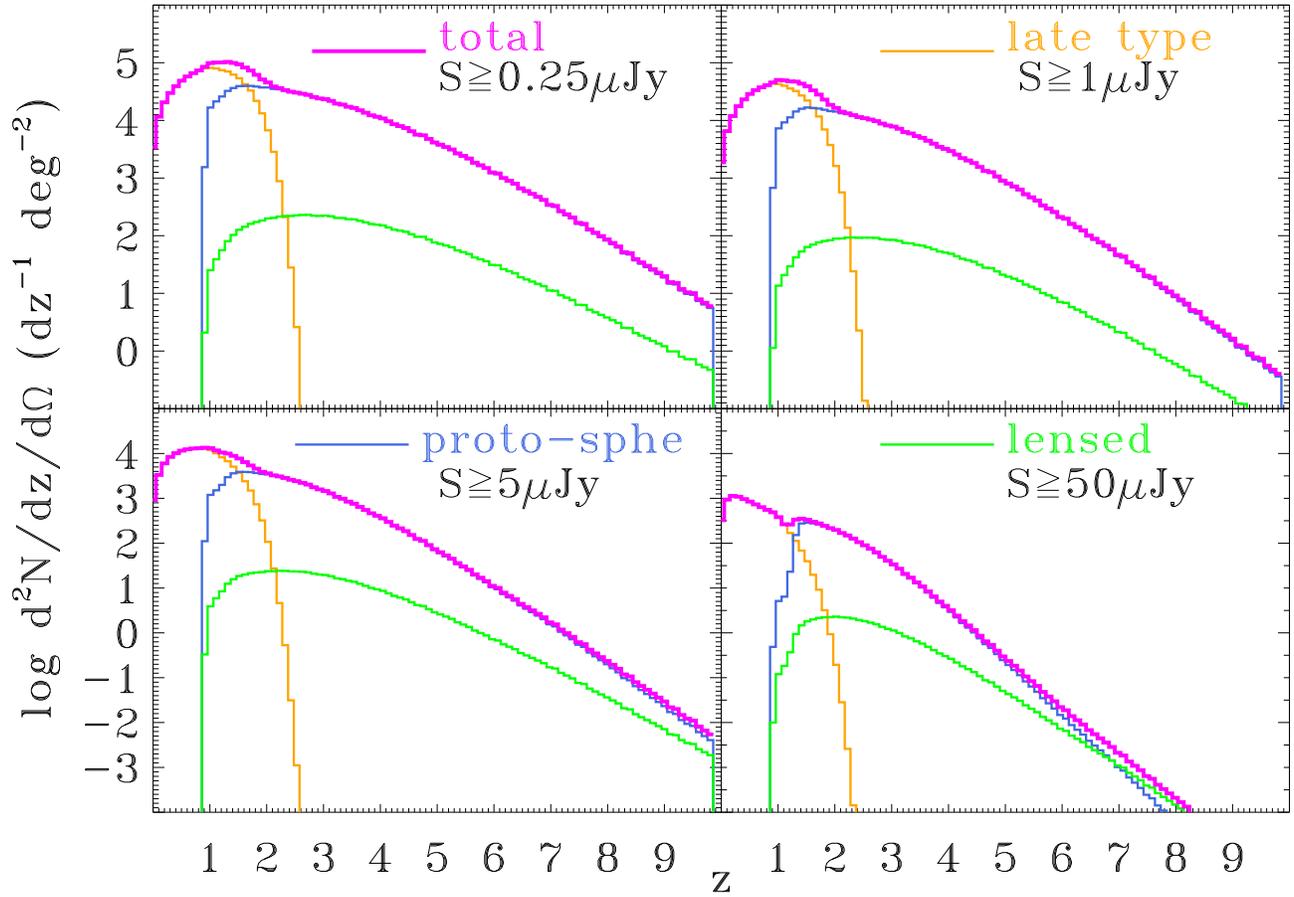}
\caption{Predicted redshift distributions at the $5\,\sigma$ detection limits
of the deepest SKA1-MID, of the MIGHTEE (MeerKAT) and of the EMU (ASKAP) 1.4
GHz surveys (see text). The lines show the contributions of the various
galaxy populations: late-type galaxies (yellow); unlensed proto-spheroidal
galaxies (blue); strongly lensed proto-spheroidal galaxies (green); the
magenta line shows the total.}
 \label{fig:RedshiftDistr}
\end{figure}

\begin{figure}
\centering
\hspace{+0.0cm}
\includegraphics[width=1\textwidth, angle=0]{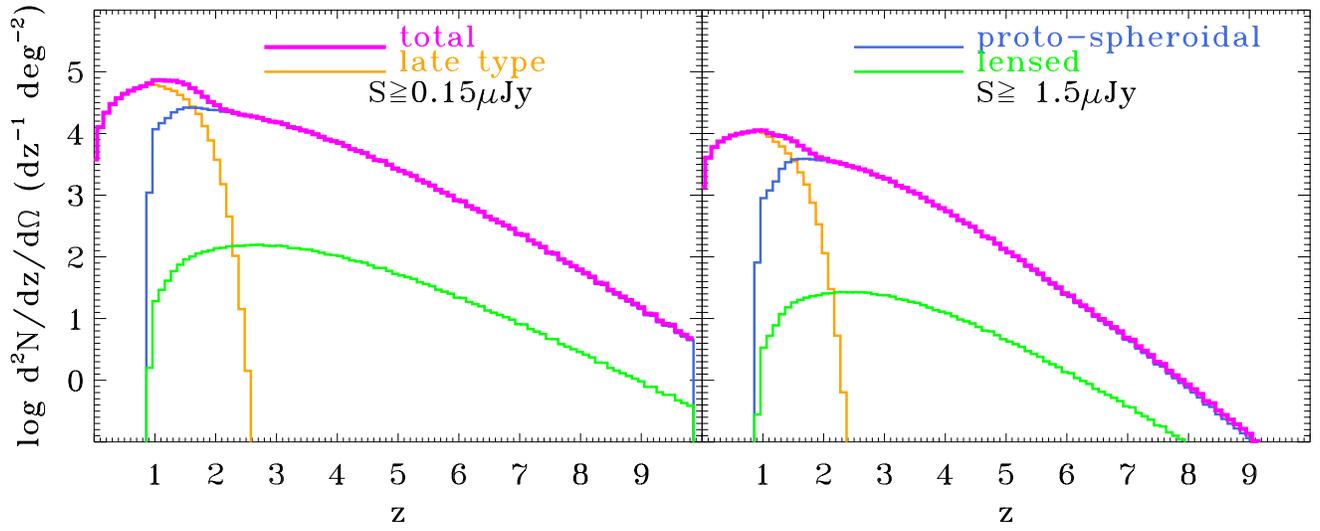}
\caption{Predicted redshift distributions at the 5$\sigma$ detection limits
of SKA1-MID 10  GHz surveys (see text). The lines show the contributions of
the various galaxy populations: late-type galaxies (yellow); unlensed
proto-spheroidal galaxies (blue); strongly lensed proto-spheroidal galaxies
(green); the magenta line shows the total.}
 \label{fig:zdistband5}
\end{figure}

\begin{figure}
\centering
\hspace{+0.0cm}
\includegraphics[width=\textwidth, angle=0]{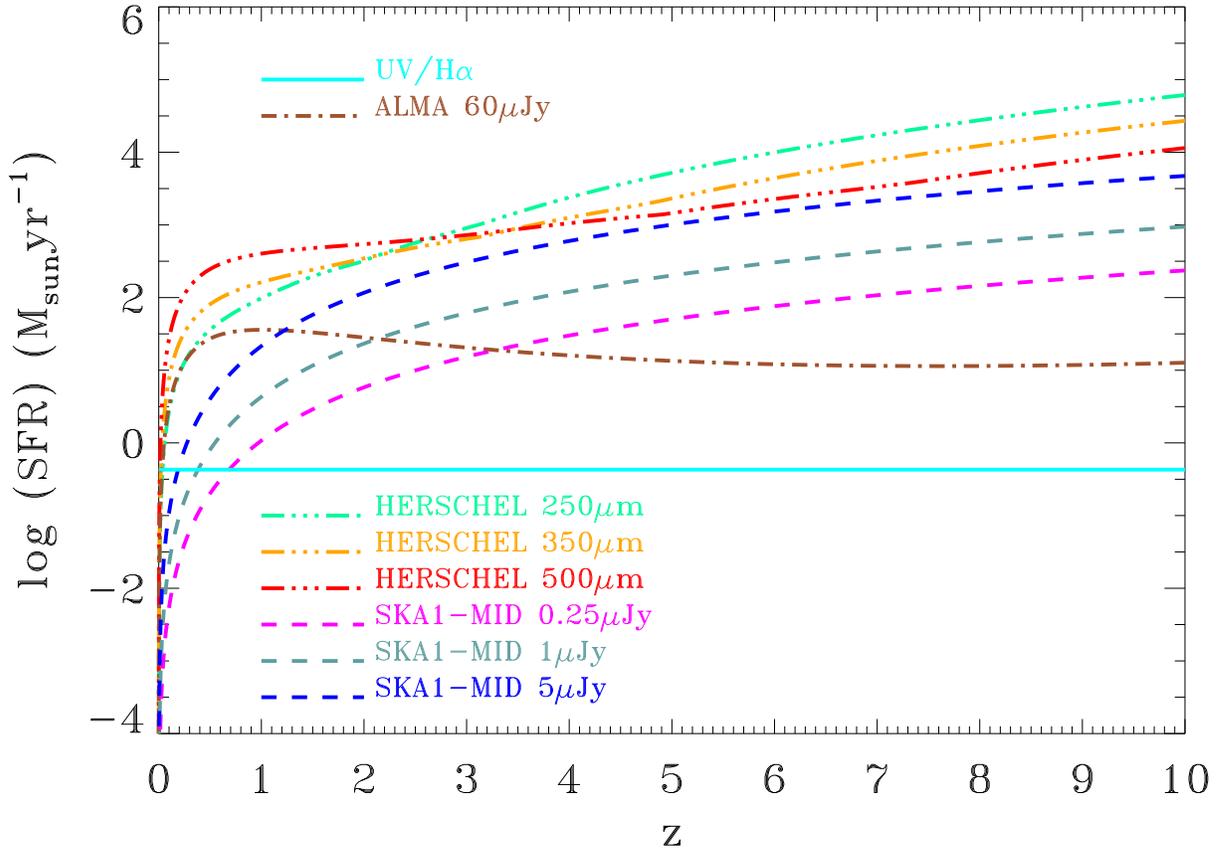}
\caption{Minimum SFR detectable by SKA1-MID surveys 1.4\,GHz down to
0.25\,$\mu$Jy (dashed magenta line), 1\,$\mu$Jy (dashed bottle green line)
and 5\,$\mu$Jy (dashed blue line), as a function of $z$, compared with the
minimum SFR detected by \textit{Herschel}/SPIRE surveys, basically
down to the confusion limits (dashed-dot-dot-dot lines, green for
250\,$\mu$m, orange for 350\,$\mu$m and red for 500\,$\mu$m) and by
UV/H$\alpha$ surveys (horizontal cyan line). The latter line corresponds to
the average minimum absolute magnitude reached by the deepest surveys, that
turns out to be almost independent of redshift. Note that, while very deep in
terms of SFR, the UV/H$\alpha$ surveys miss the dust-enshrouded star
formation. Also shown, for comparison, is the minimum SFR detectable
in the deepest ALMA maps at 1.1 and 1.3\,mm  available so far
\citep[dot-dashed brown line;][]{Carniani2015}.  }
 \label{fig:SKAcomp}
\end{figure}

The UV luminosity functions of late-type galaxies have been corrected for the
SFR dependent dust extinction and converted into SFR functions using the
\citet{KennicuttEvans2012} calibration
\begin{equation}
\log \left({L_{1500}\over L_\odot}\right) = \log \left({{\rm SFR}\over
M_\odot\,{\rm yr}^{-1}}\right) + 9.825 - 0.4 A_{1500}(\log {\rm SFR}),
\end{equation}\label{eq:UV_SFR}
via the equation
\begin{eqnarray}
\frac{d^2N(\log {\rm SFR})}{d\log {\rm SFR} dV} &=& \frac{d^2N(\log L_{1500})}{d\log L_{1500} dV} \cdot  \frac{d \log L_{1500}}{d \log {\rm SFR}} \nonumber \\
&= &\frac{d^2N(\log L_{1500})}{d\log L_{1500} dV} \cdot \left[1 - 0.4 \frac{d A_{1500}}{d\log {\rm SFR}} \right].
\end{eqnarray}
The results are shown in Fig.~\ref{fig:SFRF}. As a further test of the model
we have used the SFR functions obtained combining UV and IR data to derive
the H$\alpha$ luminosity functions at several redshifts, adopting the
\citet{KennicuttEvans2012} calibration. To compare them with observations we
have applied the SFR dependent dust attenuation of
eq.~(\ref{eq:Ha_extinction}) in the case of late-type galaxies and the
Calzetti relation between H$\alpha$ and UV extinction in the case of
proto-spheroids:
\begin{equation}\label{eq:sph_Ha_extinction}
A_{{\rm H}\alpha} = \frac{k(6563\,\angstrom)}{k(1350\,\angstrom)}
\frac{E(B-V)_{\rm gas}}{E(B-V)_{\rm ste}} A_{1350} = \frac{3.3258}{11.039}
\frac{1}{0.44} A_{1350} = 0.685 A_{1350}.
\end{equation}
As shown by Fig.~\ref{fig:LFHa}, the agreement is generally good, with a hint
that also H$\alpha$ surveys may miss galaxies with the highest SFRs,
particularly at $z\simeq 1.5$.  Figure~\ref{fig:SFRF} shows that SFR
estimates from H$\alpha$ luminosities exceed those from UV luminosities if
both are uncorrected for dust attenuation, consistent with the lower
attenuation at the H$\alpha$ compared to UV wavelengths
[eq.~(\ref{eq:sph_Ha_extinction})].

\section{Counts of strongly lensed sources}\label{appB}

In a Singular Isothermal Sphere (SIS) model the strong lensing regime (i.e.
the formation of multiple images) corresponds to $\mu_{\rm tot}\ge2$, where
$\mu_{\rm tot}$ is the {\it total} amplification, sum of the modules of the
amplifications of the two images. In fact, in a SIS model, two (and only two)
images can form in the strong lensing regime. If we indicate with $\mu_{+}$
and $\mu_{-}$ the values of the amplifications of the two images we have that
\begin{eqnarray}
\mu_{+} & = & \frac{\beta + \theta_{\rm E}}{\beta} > 0 \\
\mu_{-} & = & \frac{\beta - \theta_{\rm E}}{\beta} < 0
\end{eqnarray}
where $\beta$ is the angular separation of the background source from the
optical axis of the lens system. In the strong lensing regime
$\beta<\theta_{\rm E}$, $\theta_{\rm E}$ being the Einstein radius
\begin{eqnarray}
\theta_{\rm E} = 4\pi \left(  \frac{\sigma_{v}}{c} \right)^{2}
\frac{D_{\rm LS}}{D_{\rm S}}.
\end{eqnarray}
Here $\sigma_{v}$ is the 1D velocity dispersion of the lens, $D_{\rm S}$ is
the angular diameter distance to the source, and $D_{\rm LS}$ is the angular
diameter distance from the lens to the source. From the above equations it
follows that
\begin{eqnarray}
\mu_{\rm tot} = \mu_{+} + |\mu_{-}| = \frac{2\theta_{\rm E}}{\beta}
\end{eqnarray}
It is worth noticing that the strong lensing regime implies
$\beta\le\theta_{\rm E}$ and, therefore, $\mu_{+}\ge2$ and $|\mu_{-}|\ge0$.

The cross-section for strong lensing with total amplification higher than a
given value $\mu_{\rm tot}\ge2$ is
\begin{eqnarray}
\Sigma(\ge\mu_{\rm tot}) = \pi\frac{4\theta^{2}_{\rm E}}{\mu_{\rm tot}^{2}}.
\label{eq:Sigma_imageTOT}
\end{eqnarray}
Similarly, the cross section for strong lensing with the second image (the
fainter one) having the module of the amplification higher than a given value
$|\mu_{-}|$ is
\begin{eqnarray}
\Sigma(\ge|\mu_{-}|) = \pi\frac{\theta^{2}_{\rm E}}{(1+|\mu_{-}|)^{2}}
\label{eq:Sigma_image2}
\end{eqnarray}
Once the lensing cross section is known, the corresponding probability is
given by \citep[in the ``non-overlapping'' cross-sections approximation, i.e.
neglecting lensing by more than one foreground mass;][]{Peacock1982}
\begin{eqnarray}
P(>\mu|z_{\rm S})  = \int_{z_{\rm L}^{\rm min}}^{z_{\rm L}^{\rm max}} \int_{M_{\rm L}^{\rm min}}^{M_{\rm L}^{\rm max}} \Sigma(>\mu|M_{\rm L},z_{\rm L},z_{\rm S}) \frac{{\rm d}^2 N_{\rm L}}{{\rm d}M_{\rm L}{\rm d}z_{L}}{\rm d}M_{\rm L}{\rm d}z_{\rm L},
\label{eq:Pmu_general_formula}
\end{eqnarray}
where $z_{\rm S}$ and $z_{\rm L}$ are the redshifts of the background source
and of the lens, respectively, while ${\rm d}^2 N_{\rm L}/{\rm d}M_{\rm
L}{\rm d}z_{L}$ is the number density of lenses per unit mass and redshift
interval. The usual choice for the mass distribution of the lenses is the
mass function of the dark matter halos derived from N-body simulation
\citep[e.g.,][]{ShethTormen1999}.

The halo mass is related to $\theta_{\rm E}$ (and therefore to the the
lensing cross-section) via the 1D velocity dispersion of the lens,
$\sigma_{v}$,  by
\begin{eqnarray}
\sigma^{2}_{v} = \frac{GM_{\rm L}}{2R_{\rm L, vir}}
\end{eqnarray}
where $R_{\rm L, vir}$ is the virial radius of the lens that we compute
assuming a virialization redshift $z_{\rm L, vir}=2.5$
\citep[see][]{Lapi2012}.

Given the {\it true} number density of un-lensed sources per unit logarithmic
interval in flux density and per unit interval in redshift, $({\rm d}^2
N/{\rm d}{\log}S\,{\rm d}z)_{\rm T}$, the corresponding number density of
lensed sources is computed as
\begin{eqnarray}
\left[ \frac{{\rm d}^2N(S, z_{\rm S})}{{\rm d}{\log}S\, {\rm d}z_{\rm S}} \right]_{\rm L}
= \int_{\mu_{\rm min}}^{\mu_{\rm max}} \left[ \frac{{\rm d}^2 N(S/\mu, z_{\rm S})}{{\rm d}{\log}S \,
    {\rm d}z_{\rm S}} \right]_{\rm T} p(\mu|z_{\rm S}) {\rm d}\mu,
\label{eq:ciccio}
\end{eqnarray}
where $S$ is the {\it measured} flux density of the sources and $p(\mu|z_{\rm
S}){\rm d}\mu$ is the probability that a source at redshift $z_{\rm S}$ has
its flux density boosted by a factor $\mu$, within ${\rm d}\mu$, because of a
lensing event. This probability is obtained from
eq.~(\ref{eq:Pmu_general_formula}) as $p(\mu|z_{\rm S})=-dP(>\mu|z_{\rm
S})/d\mu$. The upper integration limit,  $\mu_{\rm max}$, is set by the size
of the background source, with $\mu_{\rm tot, max}=\infty$ for a point
source.

\begin{figure}
\centering
\includegraphics[width=1\textwidth]{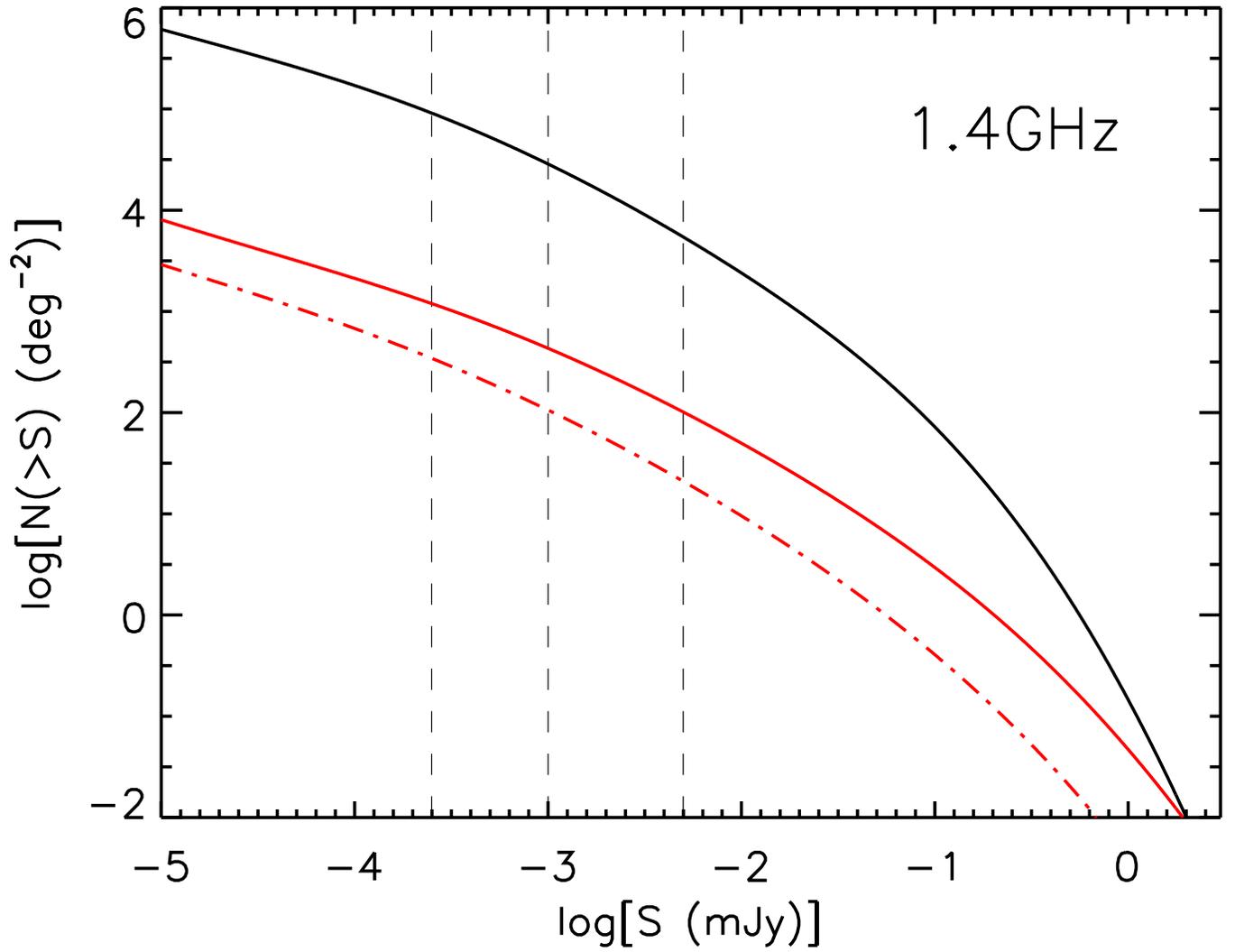}
\caption{Predicted integral number counts of proto-spheroids at 1.4 GHz. The
solid black curve represents the un-lensed counts. The red curves represent
the counts of strongly lensed galaxies as a function of the \textit{total}
(sum of all images) flux density (solid) and of the flux density of the
fainter of the two images (dot-dashed). The vertical lines represent the
three SKA1-MID 5\,$\sigma$ detection limits for unresolved sources
(i.e 0.25\,$\mu$Jy, 1\,$\mu$Jy, 5\,$\mu$Jy).}\label{fig:negrello}
\end{figure}

\begin{figure}
\centering
\hspace{+0.0cm}
\includegraphics[width=\textwidth, angle=0]{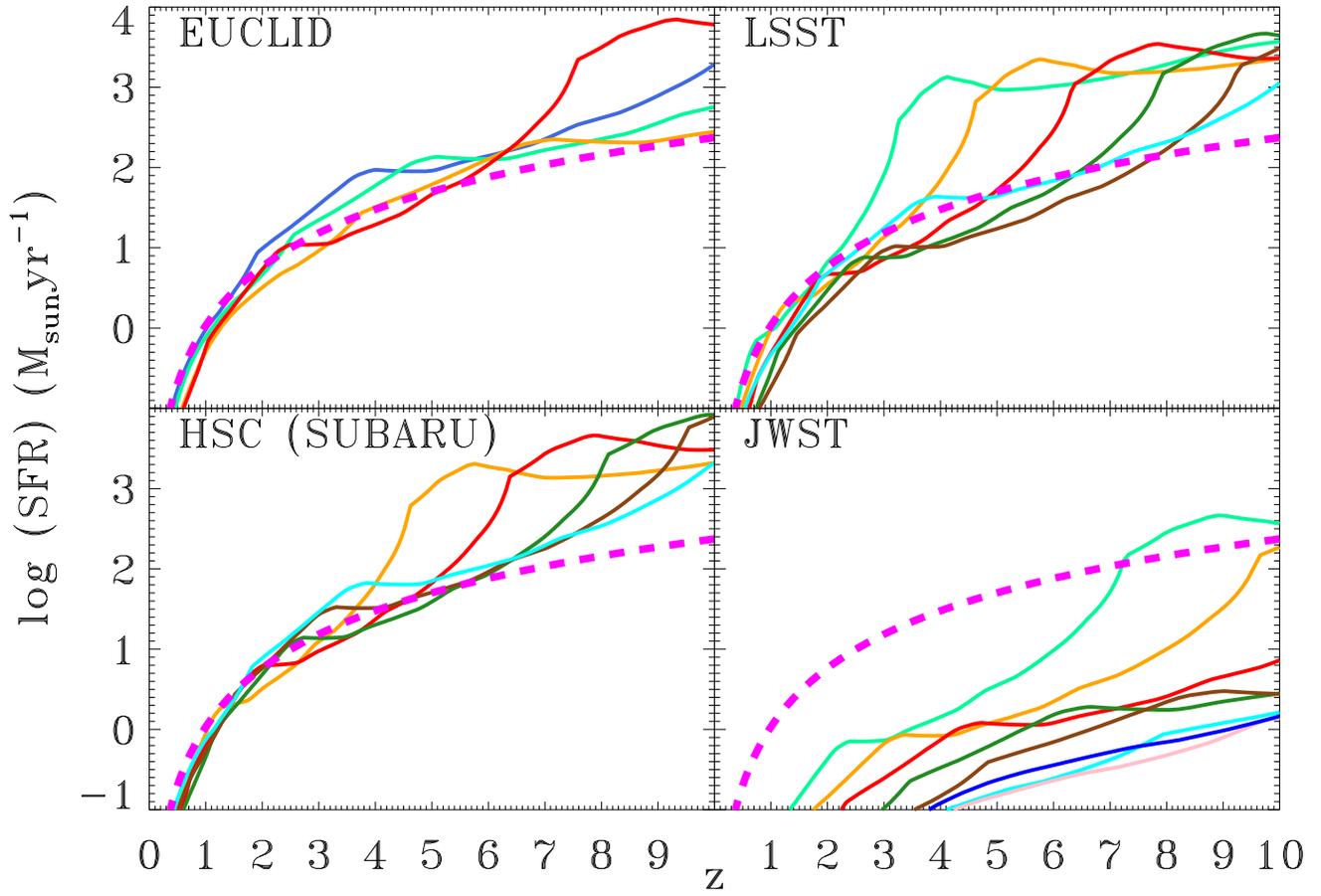}
\caption{Minimum SFR detectable by the deepest SKA1-MID survey at 1.4\,GHz
(down to 0.25\,$\mu$Jy, dashed magenta line), compared with the minimum SFR
detectable in each spectral band by the \textit{Euclid} deep survey, by the
Subaru  HSC ultra-deep survey, by the LSST deep drilling fields
and by the JWST DWS for the obscured SED shown in the right-hand panel of
Fig.~\ref{fig:SKA-JWST SED}. }\label{fig:SKAobscured}
\end{figure}

The solid and the dashed red curves in Fig.~\ref{fig:negrello} have been
derived by plugging into eq.\,(\ref{eq:Pmu_general_formula}) the lensing
cross-section given by eq.\,(\ref{eq:Sigma_imageTOT}) and
Eq.\,(\ref{eq:Sigma_image2}), respectively, and assuming in both cases
$\mu_{\rm tot, max}=30$ (translating to $|\mu_{-}|\le \mu_{\rm tot,
max}/2-1=14$).

\end{appendix}


\clearpage

\clearpage

\clearpage

\clearpage

\clearpage

\clearpage

\clearpage

\clearpage

\clearpage

\clearpage

\clearpage

\clearpage

\clearpage

\clearpage

\clearpage

\begin{figure}
\hspace{+0.0cm}
\includegraphics[width=0.5\textwidth, angle=0]{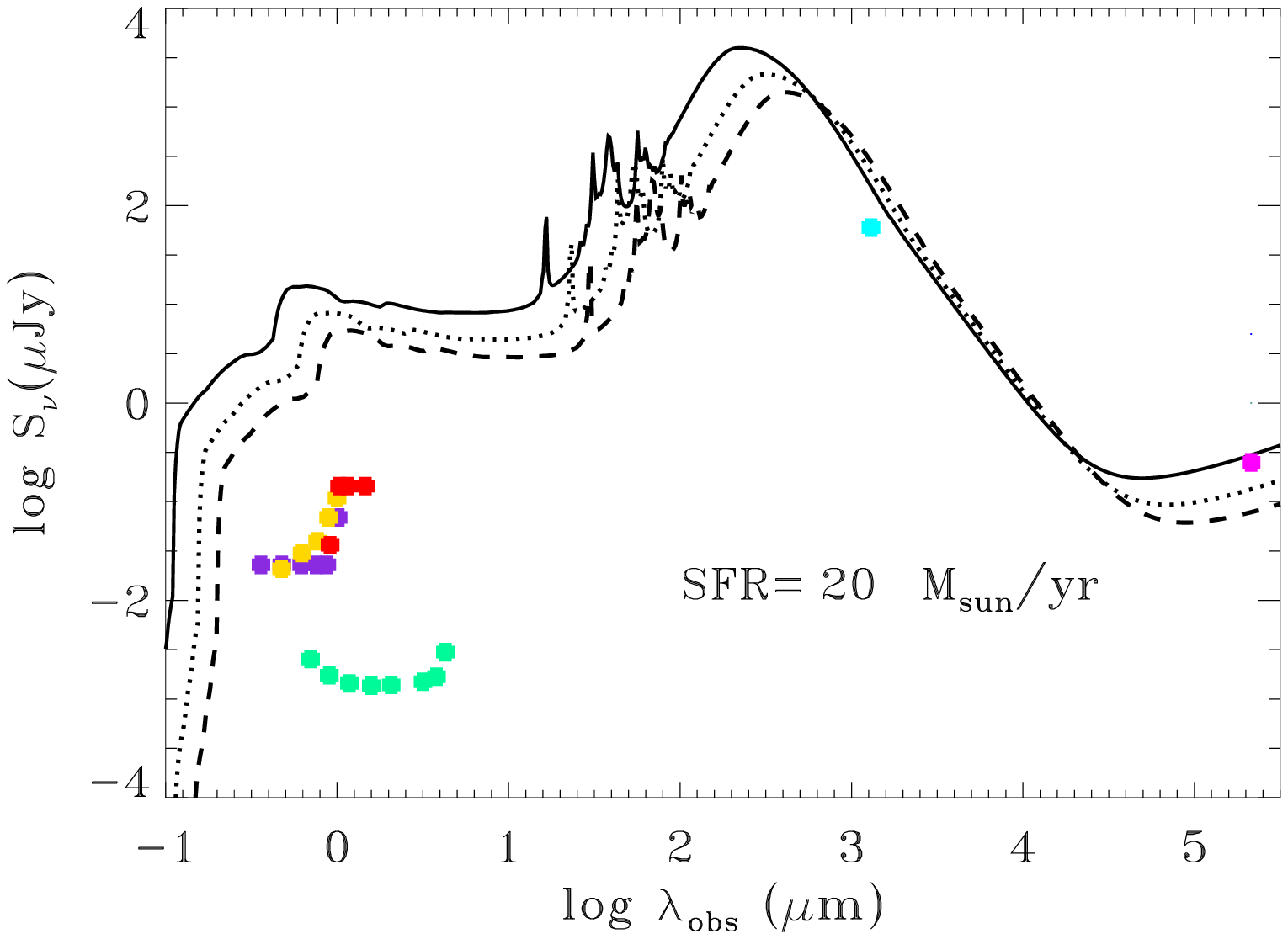}
\includegraphics[width=0.5\textwidth, angle=0]{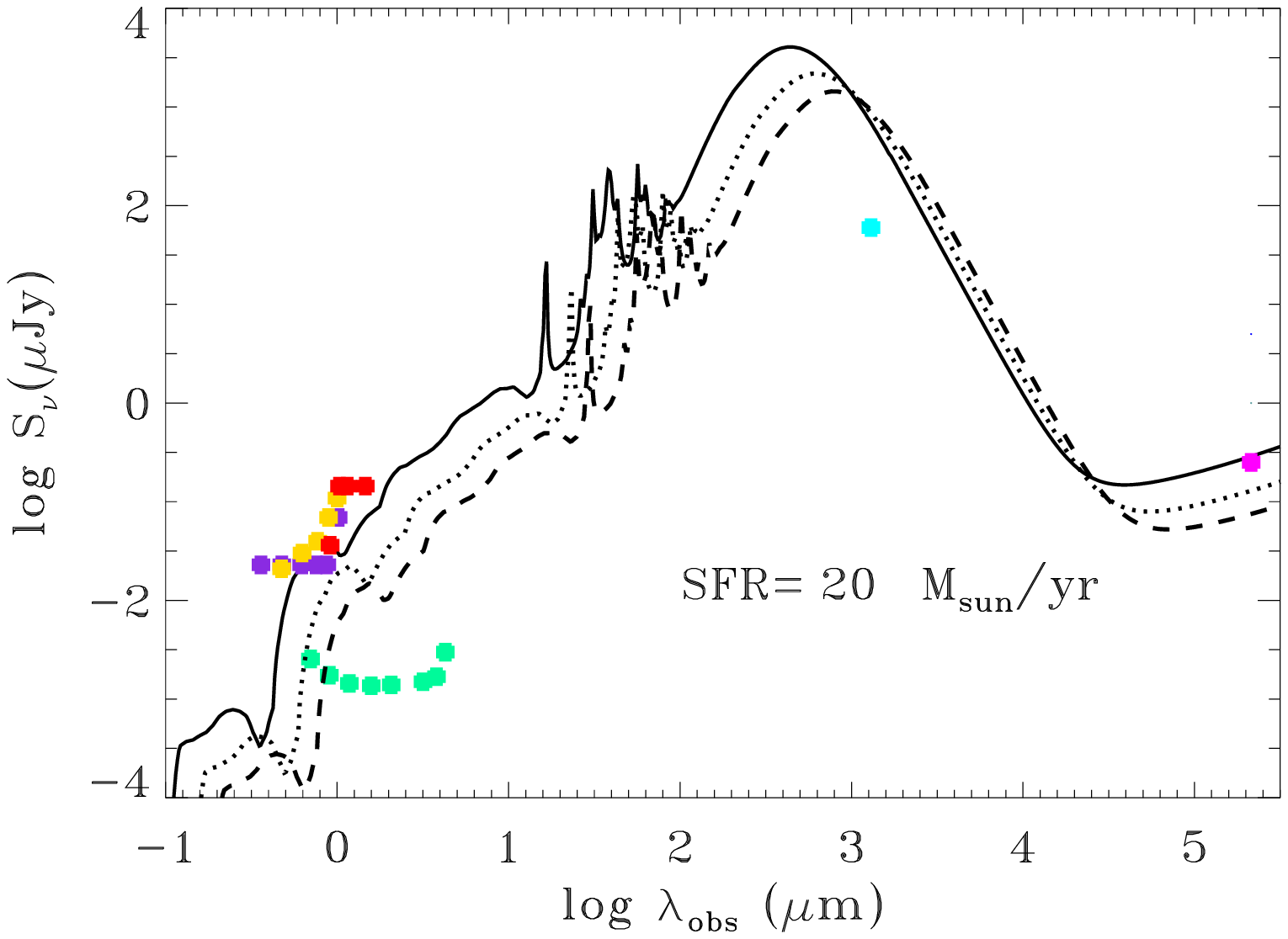}
\caption{\textit{Left panel:} Detection limits of deep optical/near-IR
surveys (left) and of the deepest SKA1-MID survey (magenta symbol on the
right) compared  with the spectral energy distribution of an UV-bright young
galaxy with a $\hbox{SFR}= 20\,M_\odot\,\hbox{yr}^{-1}$ at $z=2$ (solid), 4
(dotted) and 6 (dashed).  We have considered the following surveys (from top
to bottom): \textit{Euclid} deep (red), Subaru HSC ultra-deep
(yellow), LSST deep drilling fields (violet) and JWST (green). The
cyan point corresponds to the flux density limit of the deepest ALMA maps at
1.1 and 1.3\,mm  available so far \citep[dot-dashed brown
line;][]{Carniani2015}. The SED was computed using the GRASIL  package
\citep{Silva1998} for a \citet{Chabrier2003} Initial Mass Function (IMF), a
galaxy age of 10\,Myr and a metallicity of $0.1\,Z_\odot$. \textit{Right
panel:} same as the left panel but for an obscured SED with the same IMF and
SFR, a galactic age of 1\,Gyr and a metallicity of $3\,Z_\odot$.}
 \label{fig:SKA-JWST SED}
\end{figure}

\clearpage

\begin{figure}
\centering
\hspace{+0.0cm}
\includegraphics[width=\textwidth, angle=0]{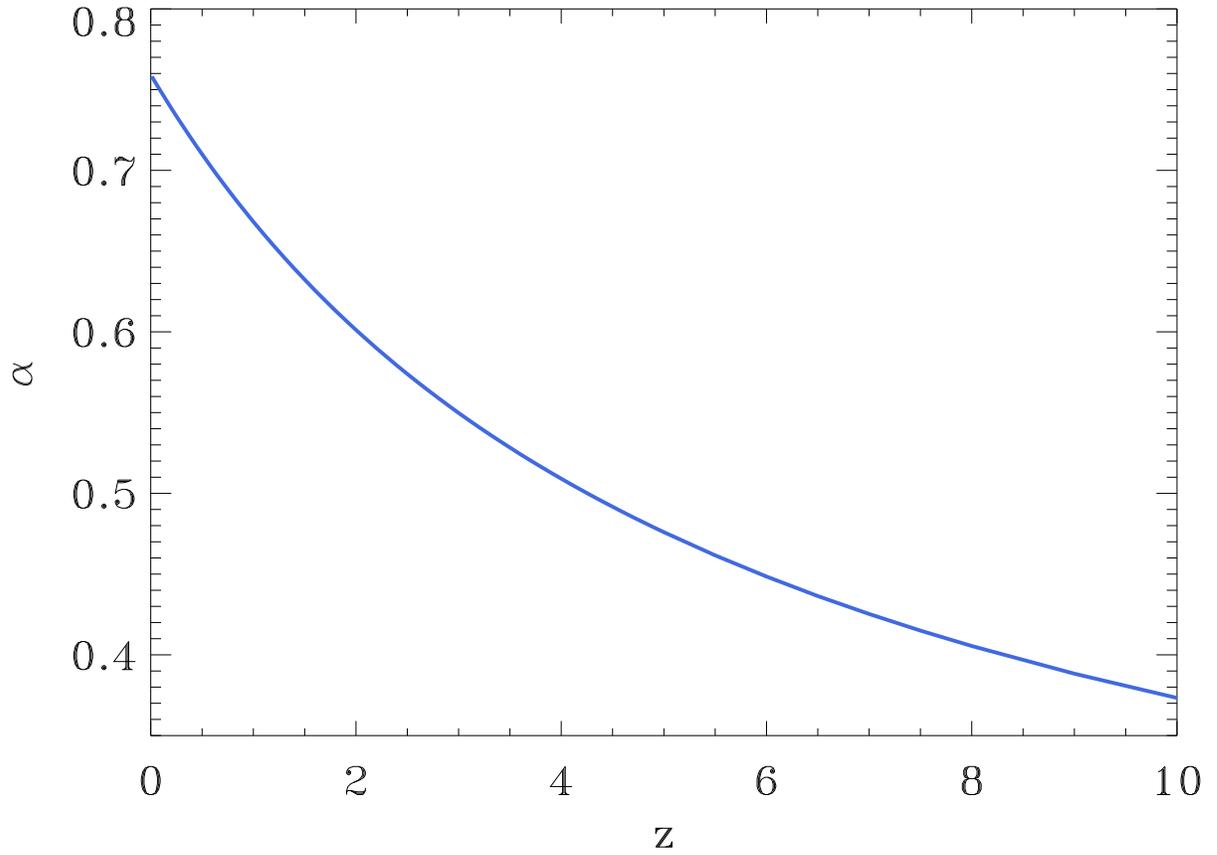}
\caption{Spectral index between 1.4 and 4.8\,GHz, $\alpha$ ($S_\nu \propto
\nu^{-\alpha}$), in the observer frame, resulting from the combination of
synchrotron and free-free emissions, as a function of $z$. The measured
spectral index is thus a rough redshift indicator. Note, however,
that the plotted curve does not take into account the possible increase of
the synchrotron suppression in galaxies with increasing redshift due to
inverse Compton scattering off the CMB photons (see
Sect.~\ref{sect:calibration}). This effect could hasten the convergence of
the mean $\alpha(z)$ towards the free-free spectral index ($\alpha\simeq
-0.1$) and introduce significant scatter especially at high $z$ since this
effect is highly dependent on how compact the star formation activity is
within a system. }
 \label{fig:alpha}
\end{figure}

\clearpage

\begin{figure}
\centering
\includegraphics[width=0.95\textwidth]{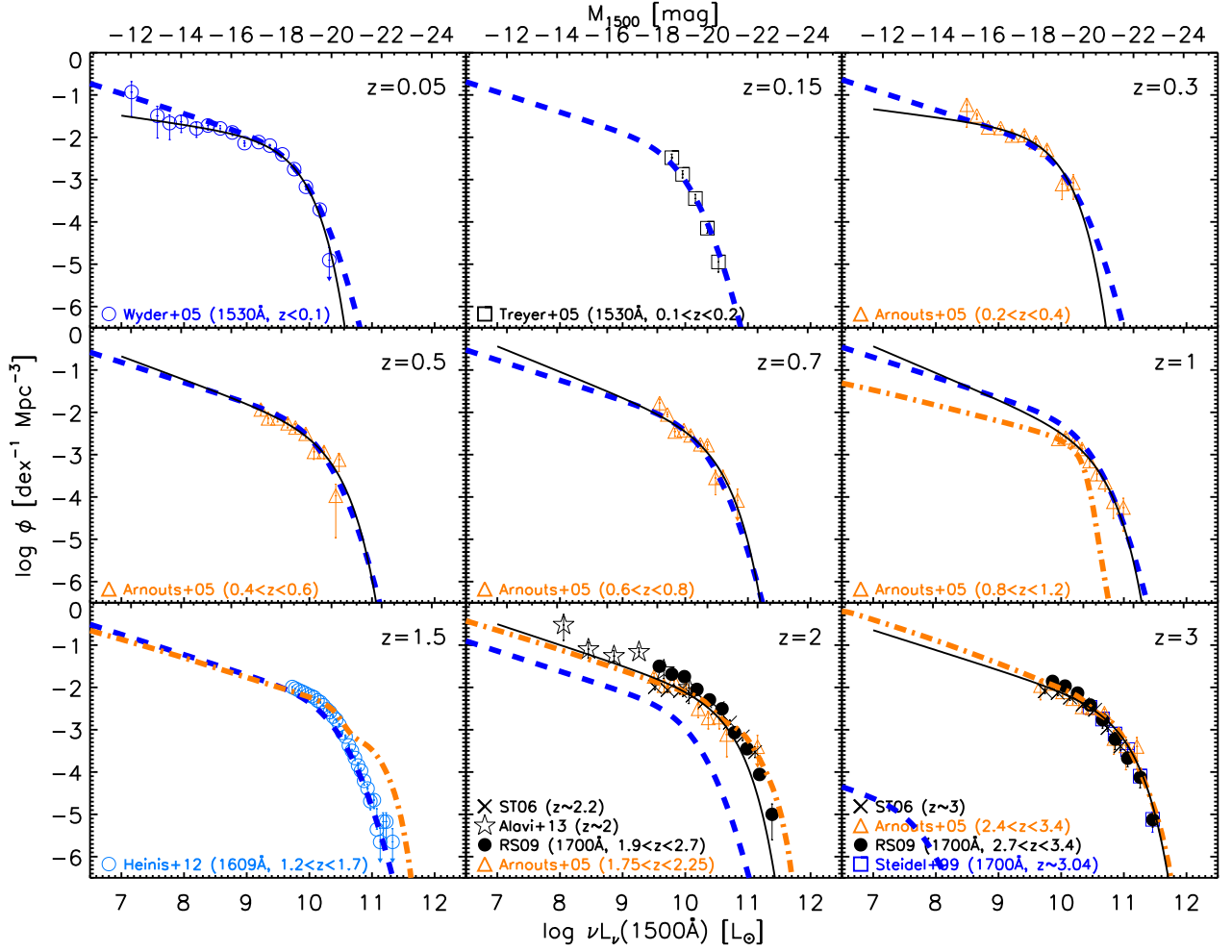}
\caption{Comparison of the observed 1500\,{\AA} luminosity functions at
several redshifts with the model described in the text. The dashed blue lines
correspond to the pure luminosity evolution model for late-type galaxies. The
contribution of proto-spheroidal galaxies (dot-dashed orange lines) begins to
show up at $z\simeq 1.5$ and becomes dominant at higher redshifts. This
contribution was obtained from the 1350\,{\AA} luminosity functions computed
by \citet{Cai2014} assuming a flat intrinsic UV spectrum [e.g.,
$L_\nu(1500\,\hbox{\AA})=L_\nu(1350\,\hbox{\AA})$] and a \citet{Calzetti2000}
extinction curve, yielding $A_{1500} = A_{1350} \times
k(1500\,\hbox{\AA})/k(1350\,\hbox{\AA}) = 0.936 A_{1350}$. The data at
different UV wavelengths have been brought to 1500\,{\AA} in an analogous
way. In some panels we also show, for comparison, the model by \citet[][thin
solid black lines]{Arnouts2005}. The data are from
\citet{Arnouts2005,Treyer2005,Wyder2005};
\citet[][ST06]{SawickiThompson2006}; \citet[][RS09]{ReddySteidel2009}; \citet{Steidel1999};
\citet{Heinis2013,Alavi2014}. }\label{fig:LFnu_1500A}
\end{figure}

\clearpage

\begin{figure}
\centering
\includegraphics[width=0.8\textwidth]{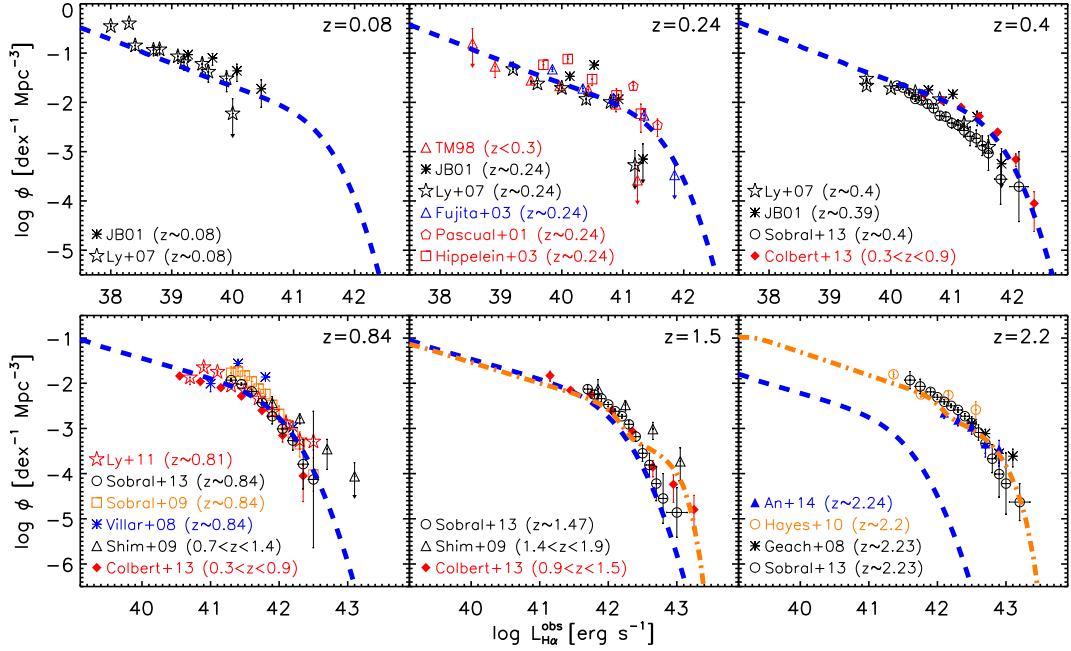}
\caption{Comparison of the observed H$\alpha$ luminosity functions not
corrected for dust attenuation with those yielded by the model, attenuated as
described in the text. The dashed blue lines are for the UV-bright late-type
galaxies and the dot-dashed orange lines for proto-spheroids.}\label{fig:LFHa}
\end{figure}

\clearpage

\begin{table}[f]
\begin{center}
\begin{tabular}{ccc}
\hline
Galaxy & redshift & $S_{95\rm GHz}\,$(mJy)\\
\hline
SJ041736-6246.8 & 0.00435 &25.56\\
SJ041959-5456.2 & 0.005017 &24.87\\
SJ044540-5914.7 & 0.00444 &21.2\\
SJ213629-5433.4 & 0.002825 &12.76\\
\hline
\end{tabular}
\end{center}
\caption{Dusty galaxies in the \citet{Mocanu2013} 95 GHz sample.}
\label{tab:dusty_gal}
\end{table}

\end{document}